\documentclass[prb,twocolumn]{revtex4}
\usepackage{graphicx}
\usepackage{hyperref}
\usepackage{amssymb}
\usepackage{dcolumn}
\usepackage{float}
\usepackage{booktabs}
\usepackage{amsmath} 
\usepackage[utf8]{inputenc}
\usepackage{lmodern}
\usepackage{color}
\definecolor{red}{rgb}{1.0,0.0,0.0}
\usepackage{makecell}
\usepackage{marginnote}

\renewcommand{\Re}{\mathrm{Re}\,}
\renewcommand{\Im}{\mathrm{Im}\,}

\newcommand{\e}{\mathrm{e}}

\renewcommand{\Re}{\mathrm{Re}\,}
\renewcommand{\Im}{\mathrm{Im}\,}

\DeclareMathAlphabet{\bi}{OML}{cmm}{b}{it}

\def\ba{\begin{aligned}}
\def\ea{\end{aligned}}
\def\be{\begin{equation}}
\def\ee{\end{equation}}
\def\bearr{\begin{eqnarray}}
\def\eearr{\end{eqnarray}}

\def\l{\left}
\def\r{\right}

\usepackage{tikz}

\begin{document}
\title{Dynamical polarization, optical conductivity and plasmon mode of a linear triple component fermionic system}
\bigskip
\author{Bashab Dey and Tarun Kanti Ghosh\\
Department of Physics, Indian Institute of Technology-Kanpur, Kanpur-208 016, India}
\begin{abstract}
We investigate the density and optical responses of a linear triple component fermionic system in both non-interacting and interacting regimes by computing its dynamical polarization function, RPA dielectric function, plasmon mode and long wavelength optical conductivity and compare the results with those of Weyl fermions and three-dimensional free electron gas. Linear triple component fermions are pseudospin-1 generalization of Weyl fermions, consisting of two linearly dispersive bands and a flat band. The presence of flat band brings about notable modifications in the response properties with respect to Weyl fermions such as induction of a new region in the particle-hole continuum, increased static polarization, reduced plasmon gap, shift in absorption edge, enhanced rate of increase in energy absorption with frequency and highly suppressed intercone transitions in the long wavelength limit. The plasmon dispersion follows the usual $\omega \sim \omega_0+ \omega_1 q^2$ nature as observed in other three-dimensional systems. 
\end{abstract}

\maketitle

\section{Introduction}
Three-dimensional (3D) semimetals having linear energy spectra around the Fermi level viz. Weyl\cite{herring,wan-weyl,xu,burkov,bulmash,murakami,halasz,smhuang,lvb,lvc,wengfang,xub} and Dirac\cite{abrikosov,wangdirac,young,murakami,murakamietal,steinberg,borisenko,liua,liub,neupane} semimetals have become breeding grounds for plethora of intriguing physical phenomena such as Fermi arc surface states\cite{wan-weyl}, chiral anomaly\cite{adler,bell,nielson}, anomalous Hall effect\cite{yang} etc. The quasiparticles close to the band-crossing nodes act as condensed-matter versions of Weyl\cite{weyl} and massless Dirac\cite{dirac} fermions theorized in high-energy physics. 
Recent studies have unveiled other classes of topological semimetals where more than two bands cross at a node and exhibit fermionic excitations with no counterpart in high energy physics\cite{bradlyn,solu,xuzhang,wieder}. It is speculated that mirror and discrete rotational symmetries in symmorphic crystals may lead to topologically protected three-fold degenerate
crossing points\cite{chang2,fulga}. First-principles calculations\cite{weng1,weng2,cheung,li} have shown that materials such as TaN,
MoP, WC, RhSi, RhGe and ZrTe can host three-band crossings in the neighborhood of the Fermi
level\cite{weng2,bqlv,he,chang3,tang,zhu}. In this paper, we deal with one such class of semimetals with three-band crossings, where quasiparticles around the nodes transform under pseudospin-1 representation\cite{takane,rao,sanchez,lvreview}. These are called triple-component semimetals\cite{bitan} (TCSs) and their low energy excitations are called triple component fermions (TCFs). The pseudospin degrees of freedom may emerge from specific admixtures of orbital and spin projections\cite{bradlyn,bradlyn2,chang}.  

The dynamics of the TCFs are governed by the Hamiltonian $H({\bf k})={\bf d}({\bf k})\cdot{\bf S}$, where ${\bf S}=(S_x,S_y,S_z)$ denote the usual spin-1 matrices and ${\bf d}({\bf k})$ is a vector function of ${\bf k}$. The band structure consists of two dispersive bands and a flat band. The TCFs can be grouped into linear, quadratic and cubic, depending on the form of ${\bf d}({\bf k})$\cite{bitan}. For linear TCFs, the energy scales linearly with all the three components of momentum. For quadratic and cubic TCFs, the energy scales linearly with $k_z$, but as $k^{2}_\perp$ and $k^{3}_\perp$ respectively in the $k_x$-$k_y$ plane, where  $k_\perp=\sqrt{k^2_x+k^2_y}$. Time-reversal symmetric TCFs arise in materials with space group symmetry 199 and 214, e.g. Pd$_3$Bi$_2$S$_2$ and Ag$_2$Se$_2$Au\cite{bradlyn}. Material realizations of time-reversal symmetry (TRS)-breaking TCFs is still absent but are predicted to be found in magnetically ordered systems\cite{bitan}.

Linear response functions serve as important tools to understand the nature of many-body correlations and excitations of a quantum system\cite{flensberg}. 
The polarizability function in momentum-frequency space obtained using the Kubo formula\cite{kubo} is known as the Lindhard function\cite{lindhard,giuliani} for conventional free electron gas (FEG) and dynamical polarization function in general. 
Its imaginary part is a measure of energy absorption by intraband or interband particle-hole excitations across the Fermi sea, which is depicted by a particle-hole continuum (PHC) in momentum-frequency space. 
The shape of PHC depends on the Fermi energy and electronic band structure of the system. The real part of the polarization function is associated with screening of the external potential. In the static limit $(\omega\to0)$, it is real valued and called the static polarization function. In one dimension, it has a logarithmic singularity  at $q=2k_F$. 
In higher dimensions, the singularities appear in its first\cite{giuliani} or second derivative\cite{minlv}. 

On inclusion of Coulomb interaction between the electrons, the dynamical polarization function gets renormalized by the dielectric function within Random Phase Approximation (RPA)\cite{rpa1,rpa2,rpa3}. 
The renormalization gives rise to a new excitation called plasmon\cite{rpa2} which is perceived as self-sustaining collective oscillation of the electrons. Plasmons have given birth to the emerging field of plasmonics\cite{plasmonics1,plasmonics2,plasmonics3}.  
In two dimensions, FEG\cite{stern}, graphene\cite{gonzalez,wangchakra,sarma,shung,ando,wunsh,pyat} and dice lattice\cite{diceplasmon}(pseudospin-1 system) host a gapless plasmon mode with dispersion $\sim \sqrt{q}\;$ at long wavelengths. In contrast, 3D FEG\cite{giuliani,mahan}, noncentrosymmetric metals\cite{sonu} and doped Weyl \cite{minlv,zhou} and Dirac semimetals\cite{sdhwang,massivedirac1,massivedirac2} exhibit gapped plasmon modes dispersing as $\sim \omega_0+ \omega_1 q^2$ in the long wavelength limit, where $\omega_0$ represents the plasmon gap. The plasmons can be probed by inelastic scattering experiments such as electron energy loss spectroscopy\cite{plasmonexp}.

The electrical conductivity in the momentum-frequency space is called the optical conductivity. It characterizes the electronic response of a material to light and is a useful tool for
extracting information about the nature of energy bands in a solid. The real part of optical conductivity is proportional to energy absorbed in the medium due to optical transitions. In the long wavelength limit ($q\to0$), the real part of optical conductivity for two-dimensional pseudospin-1/2 \cite{graphene-ando,Gusysin,nair,mak,stauber} and pseudospin-1\cite{dora,diceoptical} Dirac systems have constant universal values of $e^2/4\hbar$ and $e^2/2\hbar$ respectively beyond their absorption edges. In three-dimensions, the long-wavelength optical conductivity of Dirac and Weyl fermions linearly increases with frequency \cite{hosur,bacsi,ashby}. Similar nature of variation has been reported for higher pseudospin generalizations of Weyl fermions \cite{multifold}. Regardless of dimensions, the absorption edges of pseudospin-1/2 and pseudospin-1 fermions begin at frequencies equal to $2E_F$ and $E_F$ respectively in the zero temperature limit, where $E_F$ is the Fermi energy. 
  
The density and plasmonic responses of TCFs are still unexplored. We fill this gap in the research by making a comprehensive study of the dynamical polarization function, PHC, RPA dielectric function and plasmon mode of linear isotropic TCFs and comparing the results with those of Weyl fermions and 3D FEG.  We investigate the interplay of three bands and the effect of flat band in particular in the above responses. The flat band adds a new region in the PHC similar to the one obtained for dice lattice\cite{diceplasmon}. The static polarization function displays similar nature of variation with momentum as Weyl fermions, but with an enhanced magnitude as compared to the latter. We obtain approximate analytical expressions of dynamical polarization function and plasmon dispersion for small $q$. The plasmon mode has the usual $\sim \omega_0+ \omega_1 q^2$ dispersion as observed in other 3D electronic systems but the gap is significantly reduced with respect to Weyl fermions for the same set of parameters. This observation is supported by numerical as well as analytical results. We also obtain analytical expressions of the real part of long wavelength optical conductivity in both non-interacting and interacting regimes. In the non-interacting regime, the long-wavelength absorption edge of TCFs begins at $\hbar\omega=2E_F$ and valence-to-conduction absorption edge is absent. 
We have explained this feature using small $q$ dependence of different factors of the dynamical polarization polarization. We observe that electron-electron interactions do not affect the optical absorption edge, but reduce the magnitude of interband absorption. Apart from that, the zero frequency Drude peak vanishes and a new peak emerges at the plasmon gap.

The paper is organized as follows. In Sec. (\ref{model}), we review the low energy band structure and eigenstates of TCF. In Sec. (\ref{lindsec}), we obtain the dynamical polarization function, static polarization function and PHC of doped linear TCS. The calculation of dielectric function and plasmon modes of the system are shown in Sec. (\ref{dielsec}). A discussion on optical conductivity is presented in Sec. (\ref{opticalsection}). Finally, the results are summarized in Sec. (\ref{conc}). For the rest of the paper, `TCFs' and `Weyl fermions/semimetals' would refer to linear isotropic TCFs and isotropic type-I Weyl fermions/semimetals repectively. 

\section{Model Hamiltonian}\label{model}
The Hamiltonian of TCFs around a band touching node is given by
\begin{equation}
H({\bf k})=\hbar v_F {\bf S}\cdot {\bf k}.
\end{equation}
Here, $v_F$ is the Fermi velocity and ${\bf S}=(S_x,S_y,S_z)$ denotes the usual spin-1 matrices. The band structure comprises of three bands viz. $E_{k+}=\hbar v_F k$ (conduction band), $E_{k-}=-\hbar v_F k$ (valence band) and $E_{k0}=0$ (flat band). Denoting the pseudospin basis states $\{|s\rangle\}$ as $|\uparrow\rangle=(1\; 0\; 0)^{\mathcal{T}}$, $|0\rangle=(0\; 1\; 0)^{\mathcal{T}}$ and $|\downarrow\rangle=(0\; 0\; 1)^{\mathcal{T}}$ where $\mathcal{T}$ stands for transpose, the single-particle eigenstates $\{|\lambda({\bf k})\rangle\}$ are given by
\begin{equation}
|+({\bf k})\rangle  = \l(\begin{array}{c}
\cos^2 \frac{\theta}{2}\\
 \frac{\sin \theta}{\sqrt{2}} \e^{i \phi}\\
 \sin^2 \frac{\theta}{2} \e^{2 i \phi}	
\end{array}\r), 
\hspace{0.3cm} |-({\bf k})\rangle = \l(\begin{array}{c}
\sin^2 \frac{\theta}{2}\\
- \frac{\sin \theta}{\sqrt{2}} \e^{i \phi}\\
 \cos^2 \frac{\theta}{2} \e^{2 i \phi}	
\end{array}\r)
\end{equation}
and 
\begin{equation}
|\;0\;({\bf k})\rangle  = \l(\begin{array}{c}
- \frac{\sin \theta}{\sqrt{2}}\\
\cos \theta \e^{i \phi}\\
 \frac{\sin \theta}{\sqrt{2}} \e^{2 i \phi}	
\end{array}\r).
\end{equation}\\

where $k,\theta, \phi$ denote the usual spherical polar coordinates in momentum space.

The Hamiltonian of Weyl semimetals around one of the nodes is $H_{\text{Weyl}}({\bf k})=\hbar v_F {\boldsymbol{\sigma}}\cdot{\bf k}$ where $\boldsymbol{\sigma}=(\sigma_x,\sigma_y,\sigma_z)$  are the Pauli matrices. On the other hand, the Hamiltonian of FEG is simply $H_{\text{FEG}}=\hbar^2k^2/(2m)$ where $m$ is the effective mass.

\section{Response functions of TCFs}

\subsection{DYNAMICAL POLARIZATION FUNCTION}\label{lindsec}

A brief review of the theory of linear density response for a multi-band system is presented in Appendix(\ref{app}). The dynamical polarization function (\ref{lindhardnon}) of a non-interacting system of electrons is given by
\begin{equation}\label{lindhard}
\chi({\bf q}, \omega)=\lim_{\eta\to 0}\frac{g}{V}\sum_{{\bf k},\lambda,\lambda^\prime}
\frac{F_{\lambda,\lambda^\prime}({\bf k,k+q}) (f_{\lambda,{\bf k}} - f_{\lambda^\prime,{\bf k+q}})}{\hbar (\omega + i \eta) + E_{\lambda,{\bf k}}- E_{\lambda^\prime,{\bf k+q}}},
\end{equation}
where $g$ is the degeneracy factor, $F_{\lambda,\lambda^\prime}({\bf k,k+q})=|\langle \lambda({\bf k})|\lambda^\prime({\bf k+q})\rangle|^2$ is the overlap between the corresponding states and $f_{\lambda,{\bf k}}=[\e^{\beta(E_{\lambda,{\bf k}}-E_F)}+1]^{-1}$ is the Fermi-Dirac distribution function. 

For TCF, the interband and intraband overlaps between the dispersive bands is given by
\begin{equation}
F_{\lambda,\lambda^\prime}({\bf k,k+q})=\frac{1}{4}\l[1+\lambda\lambda^\prime \frac{{\bf k\cdot (k+q)}}{|{\bf k}||{\bf k+q}|}\r]^2, \hspace{0.5cm}\lambda,\lambda^\prime=\pm1
\end{equation}
and that between the flat and dispersive bands is 
\begin{equation}
F_{0,\lambda}({\bf k,k+q})=F_{\lambda,0}({\bf k,k+q})=\frac{1}{2}\l[1-\l(\frac{{\bf k\cdot (k+q)}}{|{\bf k}||{\bf k+q}|}\r)^2\r].
\end{equation}
At $T\to0$ K, the dynamical polarization function (\ref{lindhard}) takes the following form for $E_F>0$ (i.e. doped TCS):
\begin{equation}
\chi({\bf q},\omega)=\chi^{(+)}({\bf q},\omega)+\chi^{(0)}({\bf q},\omega)+\chi^{(-)}({\bf q},\omega),
\end{equation}
where 

\begin{widetext}
\begin{equation}\label{chi+}
\begin{aligned}
\chi^{(+)}({\bf q}, \omega)= & \lim_{\eta\to 0} \frac{g}{V}\sum_{{\bf k}}
\bigg[\frac{F_{+,+}({\bf k,k+q}) (f_{+,{\bf k}} - f_{+,{\bf k+q}})}{\hbar \omega + i \eta + E_{+,{\bf k}}- E_{+,{\bf k+q}}}
 + \frac{F_{+,0}({\bf k,k+q}) f_{+,{\bf k}}}{\hbar \omega + i \eta + E_{+,{\bf k}}-E_{0,{\bf k+q}}} - 
\frac{F_{0,+}({\bf k,k+q}) f_{+,{\bf k+q}}}{\hbar \omega + i \eta + E_{0,{\bf k}}-E_{+,{\bf k+q}}} \\
&+ \frac{F_{+,-}({\bf k,k+q}) f_{+,{\bf k}}}{\hbar \omega + i \eta + E_{+,{\bf k}}-E_{-,{\bf k+q}}} - 
\frac{F_{-,+}({\bf k,k+q}) f_{+,{\bf k+q}}}{\hbar \omega + i \eta + E_{-,{\bf k}}-E_{+,{\bf k+q}}}\bigg],
\end{aligned}
\end{equation}
\begin{equation}\label{chi0}
\chi^{(0)}({\bf q}, \omega)=\lim_{\eta\to 0} \frac{g}{V}\sum_{{\bf k}}
\bigg[\frac{F_{0,+}({\bf k,k+q}) f_{0,{\bf k}}}{\hbar \omega + i \eta + E_{0,{\bf k}}-E_{+,{\bf k+q}}} - 
\frac{F_{+,0}({\bf k,k+q}) f_{0,{\bf k+q}}}{\hbar \omega + i \eta + E_{+,{\bf k}}-E_{0,{\bf k+q}}}\bigg]
\end{equation}
and
\begin{equation}\label{chi-}
\chi^{(-)}({\bf q}, \omega)=\lim_{\eta\to 0} \frac{g}{V}\sum_{{\bf k}}
\bigg[\frac{F_{-,+}({\bf k,k+q}) f_{-,{\bf k}}}{\hbar \omega + i \eta + E_{-,{\bf k}}-E_{+,{\bf k+q}}} - 
\frac{F_{+,-}({\bf k,k+q}) f_{-,{\bf k+q}}}{\hbar \omega + i \eta + E_{+,{\bf k}}-E_{-,{\bf k+q}}}\bigg].
\end{equation}
\end{widetext}

Here, we have excluded the terms which represent the intraband transitions within flat and valence bands and the interband transitions between them. This is true only for $E_F>0$.\\
On non-dimensionalizing the quantities as $x=k/k_F$, $Q=q/k_F$, $\Omega=\lim_{\eta\to 0}\hbar(\omega+i\eta)/E_F=\lim_{\eta\to0} (\tilde{\omega}+i \hbar\eta/E_F)$, $\tilde{\omega}=\hbar\omega/E_F$ and $\tilde{\chi}^{(\lambda)}(Q,\Omega)=\chi^{(\lambda)}({\bf q},\omega)/\chi_F$ (where $E_F=\hbar v_F k_F$ and $\chi_F=g k^2_F/(4 \pi^2 \hbar v_F)$) and converting the summation into continuous integrals, equations (\ref{chi+}),(\ref{chi0}) and (\ref{chi-}) simplify as
 
\begin{widetext}
\begin{equation}\label{chi+d}
\begin{aligned}
&\tilde{\chi}^{(+)}(Q, \Omega)=\int_{0}^{1} \frac{x(\Omega+x)}{4Q}\log\l(\frac{\Omega^2+2\Omega x-Q^2+2xQ}{\Omega^2+2\Omega x-Q^2-2xQ}\r)dx
\\&+\int_{0}^{1} \frac{(\Omega+x)\l[-4Qx(\Omega^2+2\Omega x+x^2)-(Q^2-x^2)^2\log\l(\frac{(Q-x)^2}{(Q+x)^2}\r)+(-Q^2+\Omega^2+2 \Omega x+2x^2)^2\log\l(\frac{\Omega^2+2\Omega x-Q^2+2xQ}{\Omega^2+2\Omega x-Q^2-2xQ}\r)\r]}{16Qx(\Omega^2+2\Omega x+ x^2)}dx
\\&+\int_{0}^{1}\l[-x-\frac{Q^2-\Omega^2-2\Omega x-2 x^2}{4Q}\log\l(\frac{\Omega^2+2\Omega x-Q^2+2xQ}{\Omega^2+2\Omega x-Q^2-2xQ}\r)\r]dx
\\&+\int_{0}^{1}\frac{Q^2}{2(\Omega+x)}\l[\frac{Q^2+x^2}{2Q^2}+\frac{(Q^2-x^2)^2}{8Q^3x}\log\l(\frac{(Q-x)^2}{(Q+x)^2}\r)\r]dx +  \big(\Omega \leftrightarrow -\Omega\big),
\end{aligned}
\end{equation}
\begin{equation}\label{chi0d}
\begin{aligned}
&\tilde{\chi}^{(0)}(Q, \Omega)=\int_{0}^{Q}\frac{1}{8xQ}\bigg[\frac{2}{3}(3 Q^2 x+x^3)+2x(-2Q^2+\Omega^2-2x^2)+2Qx\Omega-\frac{(Q-x)^2(Q+x)^2}{\Omega}\log\l(\frac{Q+x}{Q-x}\r)
\\&\frac{(Q-\Omega-x)(Q+\Omega-x)(Q-\Omega+x)(Q+\Omega+x)}{\Omega}\log\l(\frac{x+Q-\Omega}{Q-x-\Omega}\r)\bigg]dx 
\\&+\int_{Q}^{\Lambda}\frac{1}{8xQ}\bigg[\frac{2}{3}(3 x^2 Q+Q^3)+2Q(-2Q^2+\Omega^2-2x^2)+2Qx\Omega-\frac{(Q-x)^2(Q+x)^2}{\Omega}\log\l(\frac{x+Q}{x-Q}\r)
\\&\frac{(Q-\Omega-x)(Q+\Omega-x)(Q-\Omega+x)(Q+\Omega+x)}{\Omega}\log\l(\frac{x+Q-\Omega}{x-Q-\Omega}\r)\bigg]dx + \big(\Omega \leftrightarrow -\Omega\big)
\end{aligned}
\end{equation}
and
\begin{equation}\label{chi-d}
\begin{aligned}
&\tilde{\chi}^{(-)}(Q, \Omega)=\int_{0}^{Q}\frac{1}{16xQ}\bigg[2x(2Q^2-\Omega^2+6\Omega x- 11x^2)-2Qx(\Omega -5x)-\frac{2}{3}(3Q^2x+x^3)-\frac{(Q-x)^2(Q+x)^2}{x-\Omega}\log\l(\frac{x+Q}{Q-x}\r)
\\&-\frac{(Q^2-(\Omega-2x)^2)^2}{\Omega-x}\log\l(\frac{2x+Q-\Omega}{Q-x}\r)\bigg]dx 
\\&+\int_{Q}^{\Lambda}\frac{1}{16xQ}\bigg[2Q(2Q^2-\Omega^2+6\Omega x- 11x^2)-2Qx(\Omega -5x)-\frac{2}{3}(3x^2Q+Q^3)-\frac{(Q-x)^2(Q+x)^2}{x-\Omega}\log\l(\frac{x+Q}{x-Q}\r)
\\&-\frac{(Q^2-(\Omega-2x)^2)^2}{\Omega-x}\log\l(\frac{2x+Q-\Omega}{2x-Q-\Omega}\r)\bigg]dx 
 + \big(\Omega \leftrightarrow -\Omega\big).
\end{aligned}
\end{equation}
\end{widetext}

where we have restricted the limits of integration in Eqs. (\ref{chi0d}) and (\ref{chi-d}) to an ultraviolet cutoff $\Lambda=k_c/k_F\gg1$. If the cutoff is not introduced, the upper limit of the second integral in Eqs. (\ref{chi0d}) and (\ref{chi-d}) will be infinity due to infinite bandwidth of the continuum model and the integrals will diverge.
However, the \textit{ab-initio} or tight-binding band structure of the system has a finite bandwidth. The ultraviolet cutoff incorporates this fact, although it implicitly oversimplifies the band structure by extrapolating the low-energy linear dispersion of the valence band upto the actual band minimum. The cutoff has also been used in previous works on Weyl and Dirac semimetals \cite{minlv,zhou,amit-prb,massivedirac2}. We consider $\Lambda=10$ for all the numerical results.

The dimensionless form of the dynamical polarization function is
\begin{equation} \label{chid}
\tilde{\chi}(Q, \Omega)=\tilde{\chi}^{(+)}(Q, \Omega)+\tilde{\chi}^{(0)}(Q, \Omega)+\tilde{\chi}^{(-)}(Q, \Omega).
\end{equation}

\begin{figure}
\hspace{-0.5cm}\includegraphics[trim={0cm 0cm 0cm 0.5cm},clip,width=8.6cm]{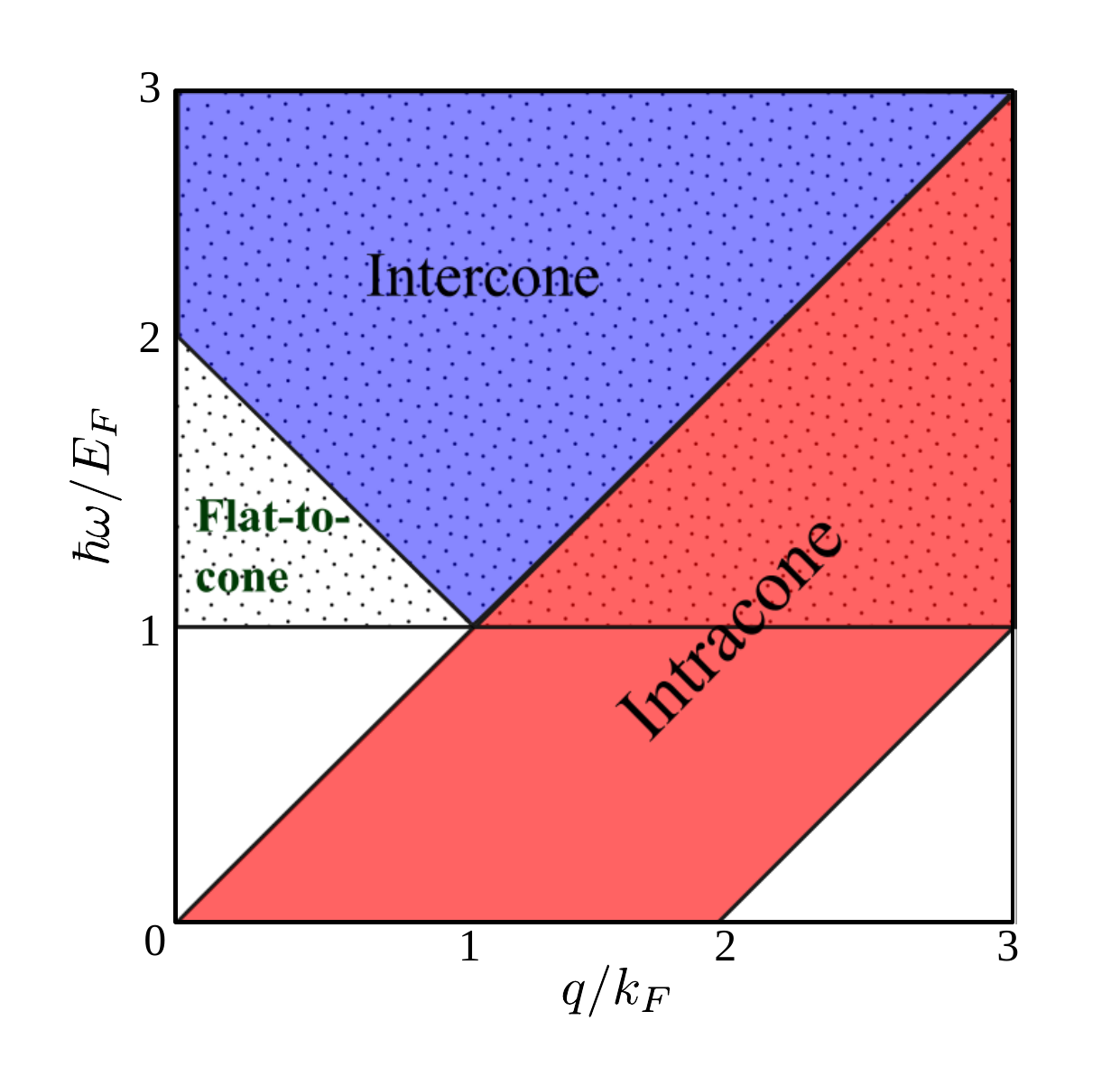}
\caption{Different regions of PHC for TCF. The dotted, violet and red regions indicates flat-to-conduction, valence-to-conduction and intra-conduction-band transitions respectively for $E_F>0$. }
\label{phcdiag}
\end{figure}

\begin{figure}
\hspace{-0.5cm}\includegraphics[trim={0cm 0cm 0cm 0.0cm},clip,width=9cm]{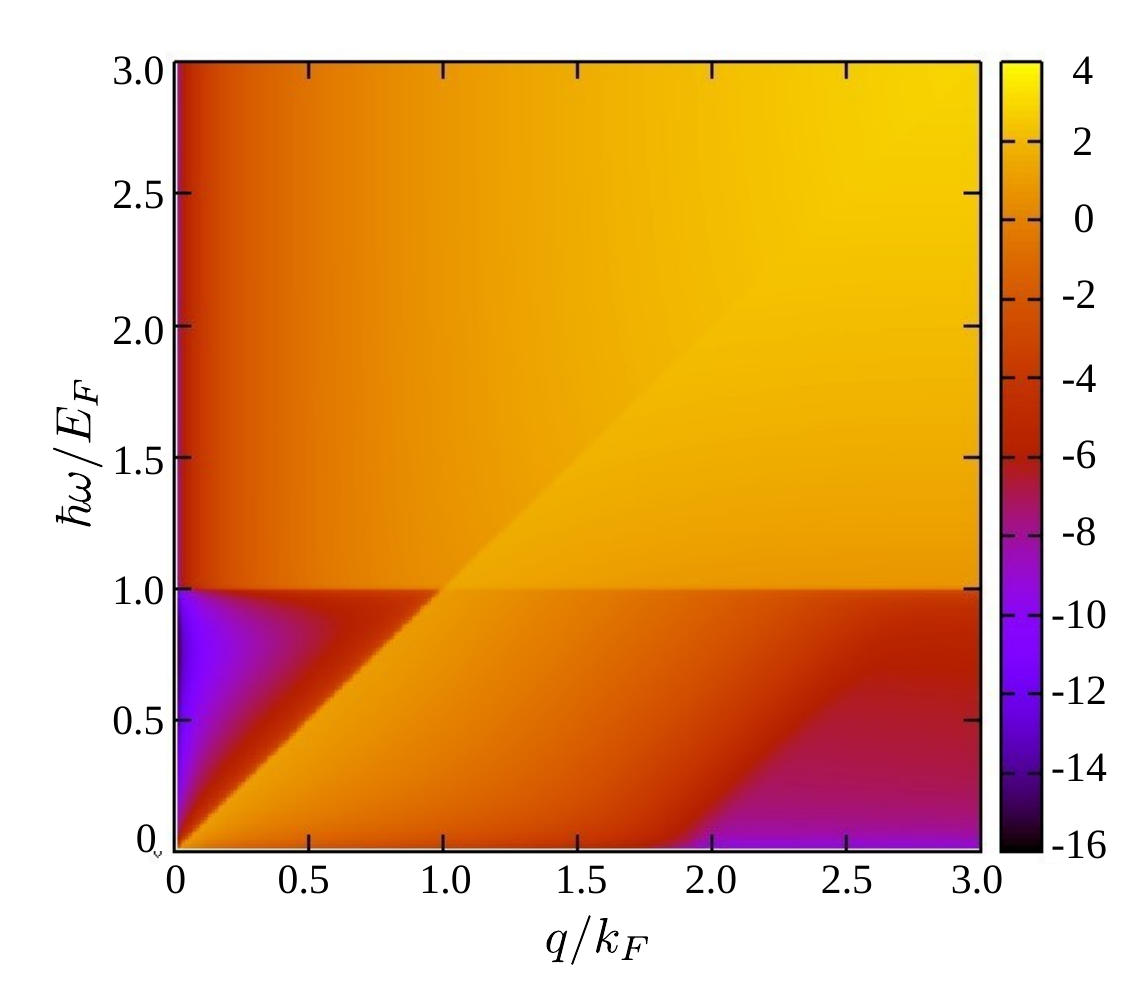}
\caption{Density plot of the natural logarithm of $\Im[\tilde{\chi}(Q,\Omega)]$ as functions of $q/k_F$ and $\hbar\omega/E_F$ for TCF.}
\label{imaglind}
\end{figure}

\begin{figure}
\includegraphics[trim={0cm 0cm 0cm 0cm},clip,width=9cm]{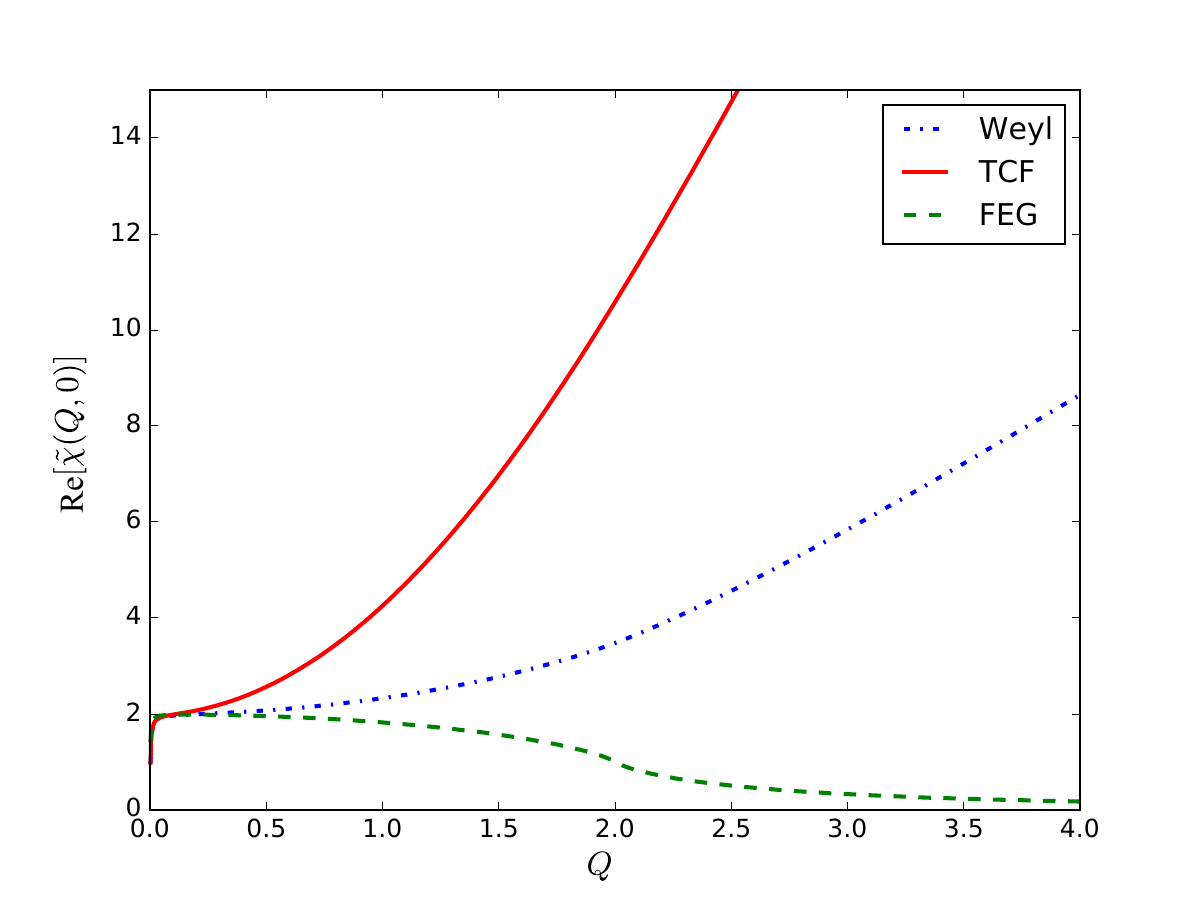}
\caption{Plots of $\Re[\tilde{\chi}(Q,0)]$ vs $Q$ for TCF, Weyl semimetal and 3D FEG. The $\Re[\tilde{\chi}(Q,0)]$ increases monotonically with $Q$ for TCFs and Weyl semimetals, but is a decreasing function of $Q$ for 3D FEG. Also, magnitude of $\Re[\tilde{\chi}(Q,0)]$ for TCFs is greater than that of Weyl semimetal for the same set of parameters. }
\label{lindhardq}
\end{figure}
 A diagram of the PHC for doped TCS ($E_F>0$) is shown in Fig.[(\ref{phcdiag})]. Like Weyl semimetals, the PHC for intraband transitions within the conduction band is bounded by $\tilde{\omega}=Q$, $\tilde{\omega}=0$ and $\tilde{\omega}=Q-2$ lines, while the interband transitions between valence and conduction bands occur in the region bounded by $\tilde{\omega}=Q$ and $\tilde{\omega}=-Q+2$ lines. The flat band introduces a new region of PHC which is absent in Weyl semimetals. The PHC for interband transitions between the flat and conduction bands is above $\tilde{\omega}=1$ line. So, the flat-to-conduction PHC overlaps those of intercone and intracone ones. These features were also observed in dice lattice\cite{diceplasmon}. The numerical plot of natural logarithm of $\Im[\tilde{\chi}(Q, \Omega)]$ as functions of $Q$ and $\tilde{\omega}$ (shown in Fig.[\ref{imaglind}]) reveals the characteristics of the PHC reasonably well albeit the sharp demarcations of different regions of absorption.

The real part of static polarization function $\Re[\tilde{\chi}(Q,0)]$ as function of $Q$ is plotted in Fig.(\ref{lindhardq}) for TCF, Weyl semimetals and 3D FEG.  In contrast to FEG, the function rises monotonically with $Q$ for both TCFs and Weyl fermions. The rise is appreciably higher in TCFs, which implies enhanced density modulation and screening as compared to Weyl fermions. This can be attributed to the presence of flat band. Moreover, the first derivative of the function is continuous at $\tilde{\omega}=2$ for both TCFs and Weyl fermions, unlike FEG.

It is cumbersome to find the analytical expression of the real part of static polarization function for linear TCFs. Although an analytical expression may be derived after tedious calculations, it is not possible to infer its nature of variation with $Q$ from the expression. The analytical forms of static polarization function for Weyl fermions and FEG has been derived in previous works\cite{minlv,mahan}. For small $Q$, the real part of static polarization function increases as $1+Q^2/6 \log(\Lambda/2)$ for Weyl semimetals and decreases as $1-Q^2/12$ for 3D FEG. \\

\begin{figure}
\hspace{-0.5cm}\includegraphics[trim={0cm 0cm 0cm 0cm},clip,width=9cm]{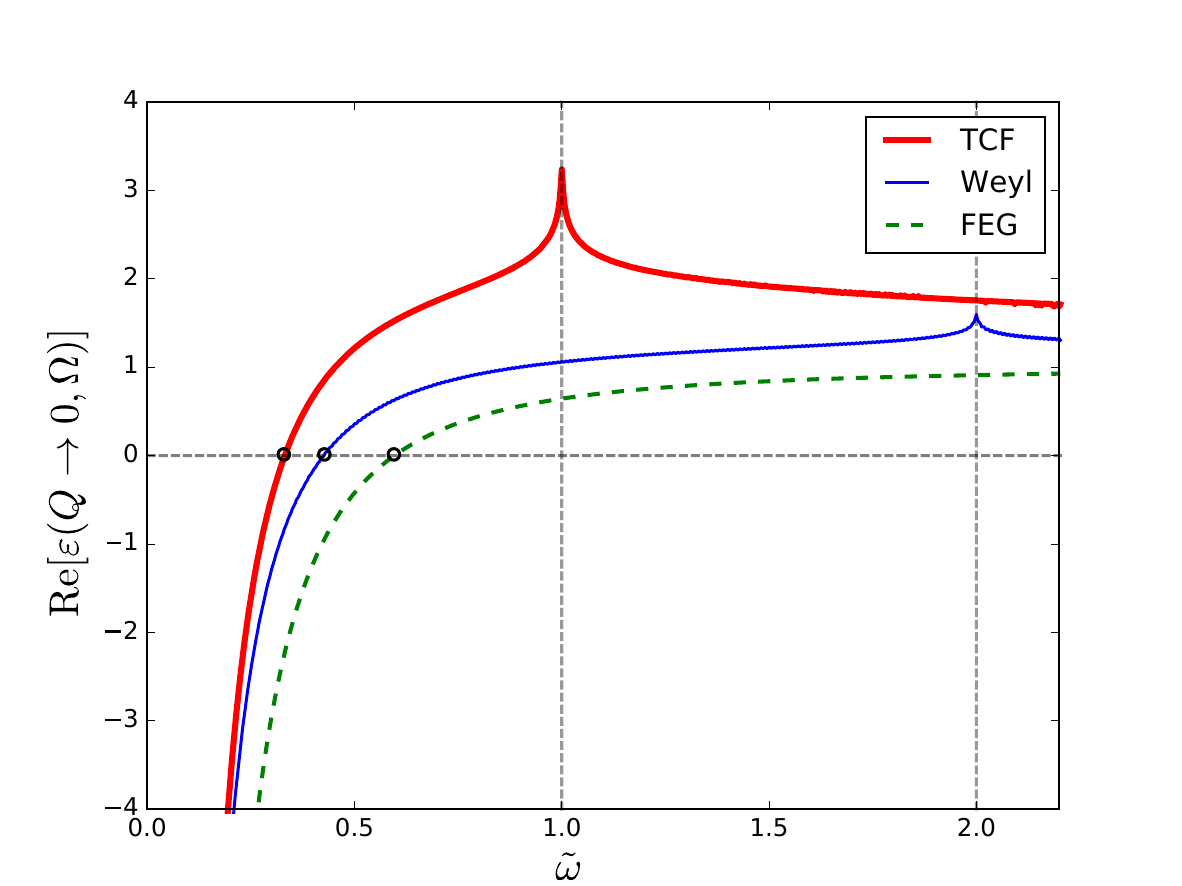}
\caption{Plots of real part of dielectric function $\Re[\varepsilon(0,\Omega)]]$ vs $\tilde{\omega}$ for TCF, Weyl semimetal and FEG. The $\Re[\varepsilon(0,\Omega)]]$ vanishes at plasmon frequencies $\tilde{\omega}^{(0)}_p$ (marked by small circles) of the respective systems. They are peaked at $\tilde{\omega}=1$ and $\tilde{\omega}=2$ for TCFs and Weyl semimetals respectively.  }
\label{epsilonomega}
\end{figure}

\begin{figure}
\hspace{-0.5cm}\includegraphics[trim={0cm 0cm 0cm 0cm},clip,width=9cm]{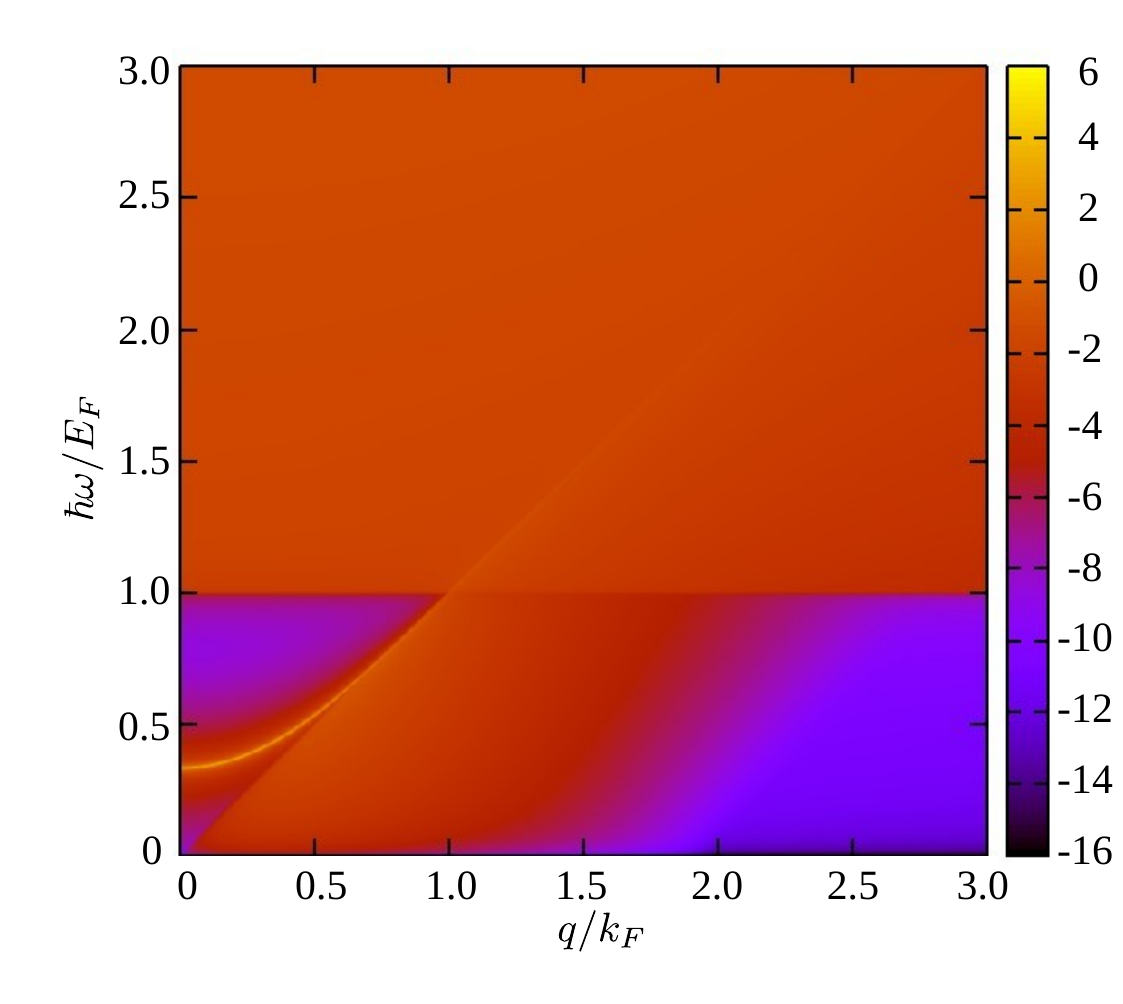}
\caption{Density plot of the natural logarithm of loss function (\ref{loss}) as a function of $q/k_F$ and $\hbar\omega/E_F$. The plasmon mode appears as bright curve in the region where ${\Im(\tilde{\chi})}$ vanishes. Hence, the mode is undamped. It continues to extend into the PHC where it gets damped into particle-hole excitations. }
\label{lossfig}
\end{figure}

\begin{figure}
\hspace{-0.5cm}\includegraphics[trim={0cm 0cm 0cm 0.0cm},clip,width=9cm]{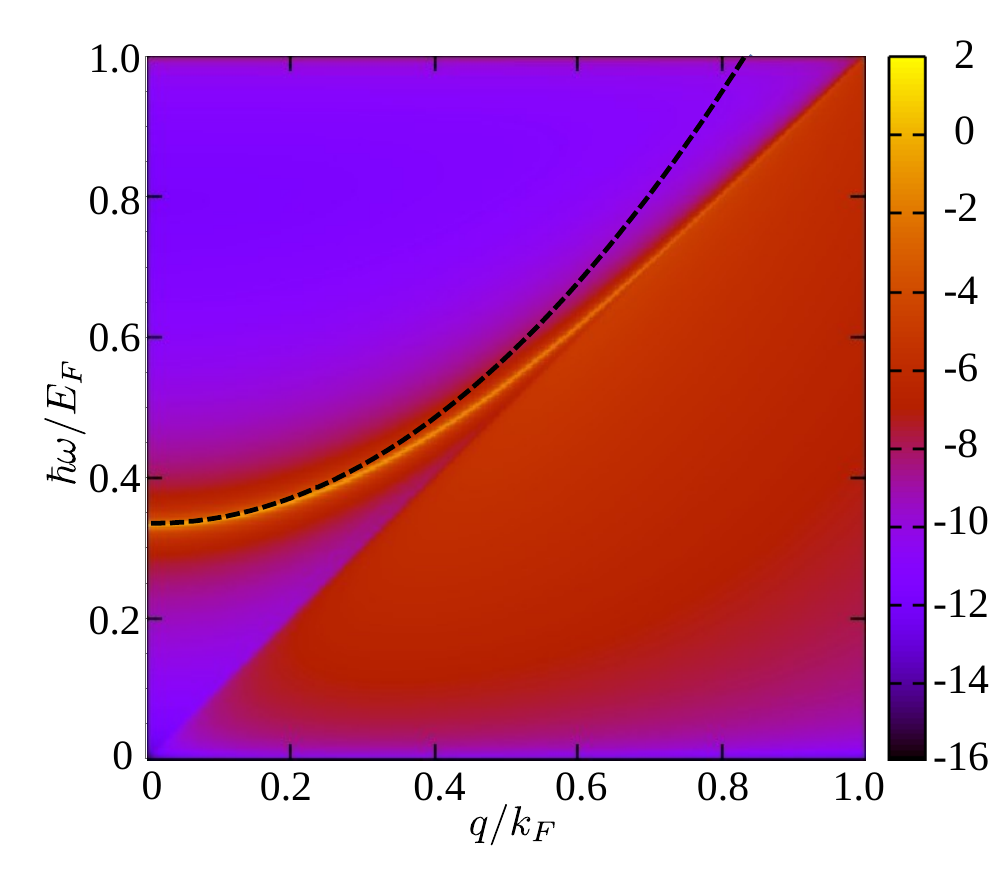}
\caption{Comparison between analytical solution of plasmon dispersion (dotted curve) for long wavelength ($Q\ll1$) regime given by Eq.(\ref{dispersion}) and numerically obtained plasmon mode in the loss function plot. The agreement is good for low $Q$ as expected.}
\label{plasmonfig}
\end{figure}

\begin{figure}
\hspace{-0.5cm}\includegraphics[trim={0cm 0cm 0cm 0.0cm},clip,width=9cm]{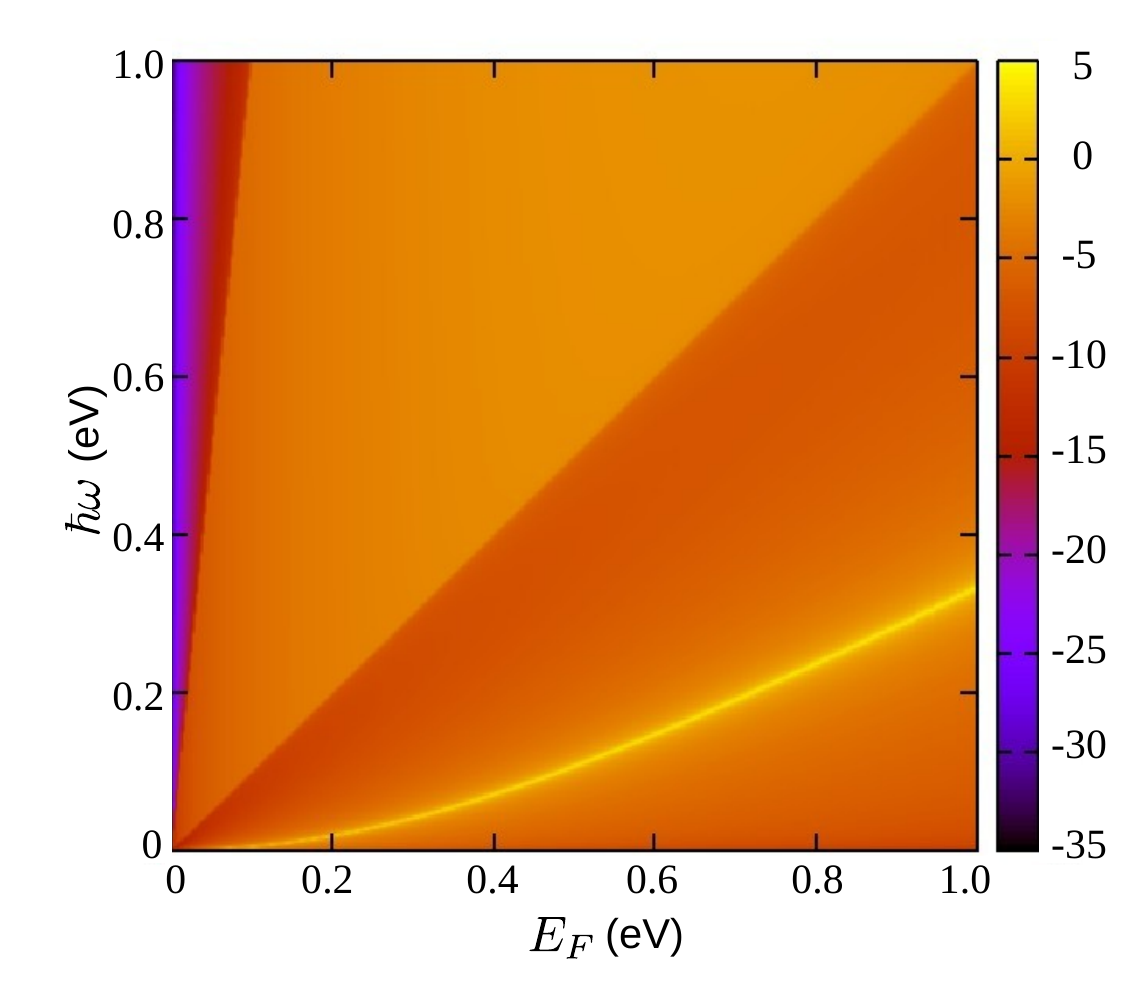}
\caption{Density plot of the natural logarithm of loss function as functions of $E_F$ and $\hbar\omega$ for TCFs at very small wavelengths $Q\to 0$ . The plasmon mode (bright yellow curve) remains undamped and its frequency increases with $E_F$.}
\label{omegaef}
\end{figure}

\begin{figure}
\hspace{-0.5cm}\includegraphics[trim={0cm 0cm 0cm 0.0cm},clip,width=9cm]{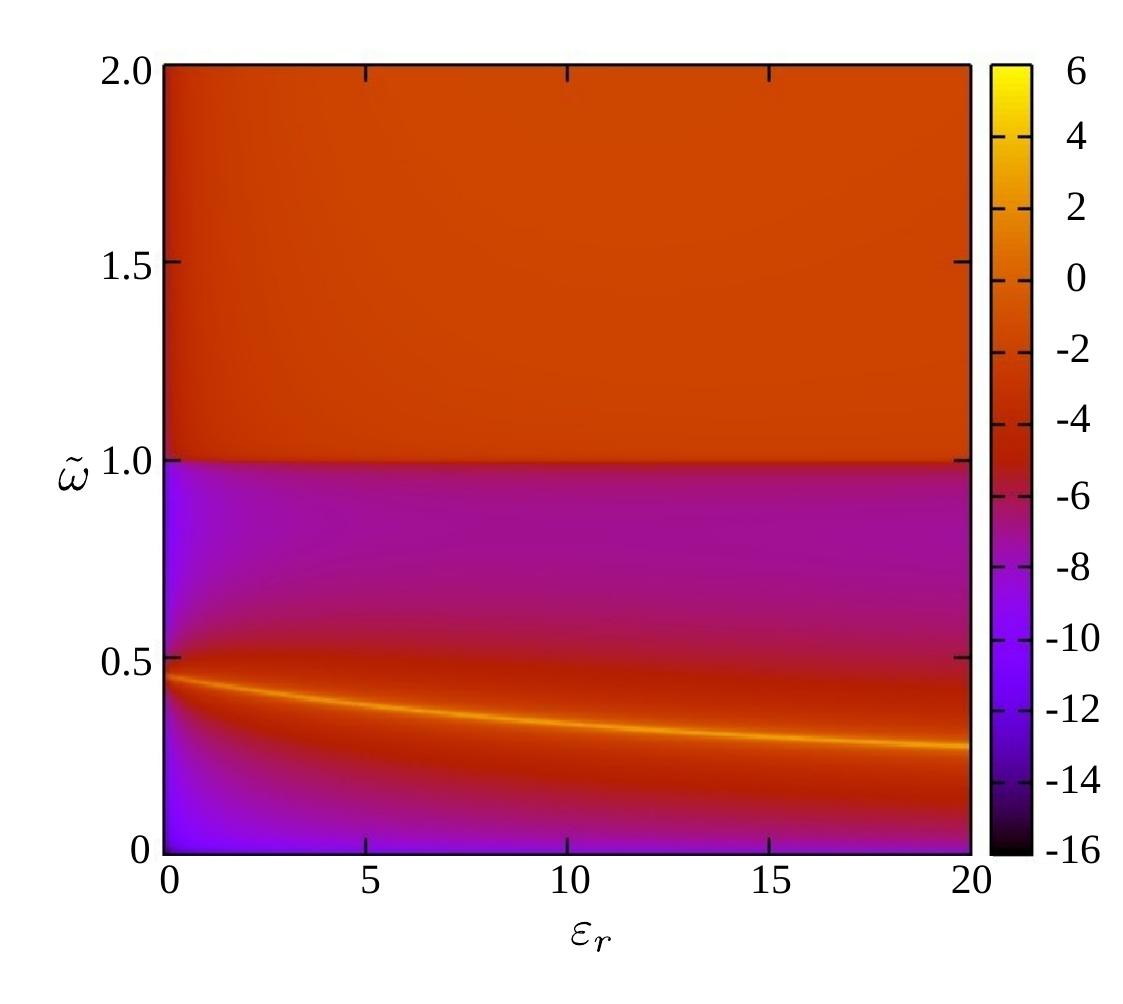}
\caption{Density plot of the natural logarithm of loss function as functions of $\varepsilon_r$ and $\tilde{\omega}$ for TCFs for very small wavelengths $Q\ll1$ . The plasmon mode (bright yellow curve) remains undamped and its frequency decreases with $\epsilon_r$.}
\label{omegaer}
\end{figure}

\subsection{DIELECTRIC FUNCTION AND PLASMONS}\label{dielsec}

For TCFs, the dielectric function (\ref{diel}) can be written as
\begin{equation}
\varepsilon(Q,\Omega)=1-\frac{C}{Q^2} \tilde{\chi}(Q, \Omega),
\end{equation} 
where $C=e^2g/(4\varepsilon_r\varepsilon_0\pi^2\hbar v_F)$. 
The undamped plasmon modes $\tilde{\omega}_p$ for TCFs can be obtained by solving the following equation for $\Omega$ and $Q$:
\begin{equation}\label{zeroes}
1-\frac{C}{Q^2} {\Re [\tilde{\chi}(Q, \Omega_p)]}=0.
\end{equation}
Since the exact solution of the Eq.(\ref{zeroes}) cannot be obtained analytically, we deduce an approximate expression of long wavelength ($Q\ll1$) and low frequency ($\tilde{\omega}\ll1$) plasmon mode of this system using the expansion of (\ref{chid}) in orders of $Q$. The dynamical polarization function for small $Q$ can be written as
\begin{equation}
\begin{aligned}\label{chiexpand}
&\tilde{\chi}(Q, \Omega)=\tilde{\chi}_{cc}(Q,\Omega)+\tilde{\chi}_{fc}(Q,\Omega)+\tilde{\chi}_{vc}(Q,\Omega),
\end{aligned}
\end{equation}
where $\tilde{\chi}_{cc}(Q,\Omega), \tilde{\chi}_{fc}(Q,\Omega)$ and $\tilde{\chi}_{vc}(Q,\Omega)$ are intra-conduction band, flat-to-conduction and valence-to-conduction (intercone) contributions respectively, given by
\begin{equation}\label{chicc}
\tilde{\chi}_{cc}(Q,\Omega)=\l(\frac{2}{3\Omega^2}Q^2+\frac{2}{5\Omega^4}Q^4+\mathcal{O}(Q^6)\r),
\end{equation}
\begin{equation}\label{chifc}
\begin{aligned}
\tilde{\chi}_{fc}&(Q,\Omega)=\int_{0}^{1}\l[\frac{-4 x}{3(\Omega^2-x^2)}Q^2+\frac{4}{15x(\Omega^2-x^2)}Q^4\r]dx\\
&+\int_{0}^{\Lambda}\l[\frac{(4x) }{3(\Omega^2-x^2)}Q^2 + \frac{4 x^2(\Omega^4-5\Omega^2 x^2)}{15(-\Omega^2 x+x^3)^3}Q^4\r]dx,
\end{aligned}
\end{equation}
and
\begin{equation}\label{chivc}
\begin{aligned}
\tilde{\chi}_{vc}(Q,\Omega)=\bigg[\int_{0}^{1}&\frac{4 x^2}{-15\Omega^2 x^3+60x^5}+
\\&\int_{0}^{\Lambda}\frac{4 x^2}{15\Omega^2 x^3-60x^5}\bigg]dx \ Q^4.
\end{aligned}
\end{equation}

Firstly, we obtain the plasmon energy gap $\tilde{\omega}^{(0)}_p=\tilde{\omega}_p(Q\to0)$ by substituting the real part of Eq.(\ref{chiexpand}) upto order of $Q^2$ in Eq.(\ref{zeroes}). The simplified form of Eq.(\ref{chiexpand}) containing only the term proportional to $Q^2$ can be written as
\begin{equation}\label{lindqsq}
\tilde{\chi}(Q^2, \Omega)=\frac{2}{3}\l(\frac{1}{\Omega^2}+\log\l[\frac{1-\Omega^2}{\Lambda^2-\Omega^2}\r]\r)Q^2.
\end{equation}
The $Q^2/\Omega^2$ term and the log term in the above expression come from the intra-conduction band and flat-to-conduction band contributions respectively. This can be seen from Eqs. (\ref{chicc}) and (\ref{chifc}). The first term of Eq. (\ref{chicc}) is clearly $\propto
Q^2/\Omega^2$ while the first and third terms of Eq. (\ref{chifc}) give the log term after integration.
Substituting the real part of Eq.(\ref{lindqsq}) in Eq.(\ref{zeroes}) gives
\begin{equation}
\begin{aligned}
&1-\frac{2}{3}C\bigg[\frac{1}{(\tilde{\omega}^{(0)}_p)^2}+\log\l|\frac{1}{\Lambda^2}\r|+\\
&\l(-1+\frac{1}{\Lambda^2}\r)(\tilde{\omega}^{(0)}_p)^2+\mathcal{O}((\tilde{\omega}^{(0)}_p)^3)\bigg]=0.
\end{aligned}
\end{equation}
Considering $\tilde{\omega}^{(0)}_p\ll1$ i.e $\hbar{\omega}^{(0)}_p\ll E_F$, we neglect the terms of the order of $(\tilde{\omega}^{(0)}_p)^2$ and higher in the above equation to get the plasmon gap as
\begin{equation}
\tilde{\omega}^{(0)}_p=\sqrt{\frac{\frac{2}{3}C}{1+\frac{2}{3}C \log \Lambda^2}}.
\end{equation}
The plasmon gap depends on the cut-off $\Lambda$. 
In terms of $E_F$, we have
\begin{equation}\label{TCFgap}
\omega^{(0)}_p=\frac{E_F}{\hbar}\sqrt{\frac{\frac{2}{3}C}{1+\frac{2}{3}C \log \Lambda^2}}.
\end{equation}
So, plasmon gap is linearly proportional to $E_F$ for large values of $E_F$. The expressions of plasmon gaps for Weyl semimetals and FEG are 
\begin{equation}\label{weylgap}
[\omega^{(0)}_p]_{\text{Weyl}}=\frac{E_F}{\hbar}\sqrt{\frac{\frac{2}{3}C}{1+\frac{1}{6}C \log \Lambda^2}}
\end{equation}
and
\begin{equation}\label{feggap}
[\omega^{(0)}_p]_{\text{FEG}}=\frac{E_F}{\hbar}\sqrt{\frac{4C^\prime}{3}},
\end{equation}
where the expression of $C$ is same as that for TCFs and $C^\prime = e^2g k_F/(4\varepsilon_r\varepsilon_0\pi^2 E_F)$ for FEG. We can see that TCFs and Weyl semimetals have similar expressions of plasmon gaps with a log term appearing in their denominators, which is absent in the expression for 3D FEG. As mentioned earlier, the log term comes from the interband transitions and it dampens the gap. The dampening is higher in TCFs than in Weyl semimetals, which can be inferred from the coefficient of the log terms.  The numerator, representing the intraband contribution, is identical for TCFs and Weyl semimetals despite the distinctive nature of their intraband overlaps. For $\hbar\omega_p\ll E_F$, the plasmon gaps for TCFs and Weyl fermions are linearly proportional to $E_F$, while that for 3D FEG is proportional to $E_F^{3/4}$. 

The variation of $\Re[\varepsilon(Q\to 0,\Omega)]$ with $\tilde{\omega}$ is shown in Fig.(\ref{epsilonomega}) for TCF, Weyl semimetals and FEG. The values of the parameters for TCFs are -- $v_F= 4 \times 10^5$ m/s, $\varepsilon_r=10$ and $g=2$ (valley degeneracy) which gives $C\approx0.347$ in Eq. (\ref{TCFgap}). The parameters for Weyl fermions are considered same as those of TCFs. For FEG, we consider the typical values of $E_F$ and $k_F$ found in metalic systems which are given by
\begin{equation}
E_F=\frac{50.1}{(r_s/a_0)^2}\text{ eV},~~\text{and}~~k_F=\frac{3.63\times 10^{10}}{r_s/a_0}~\text{m}^{-1}.
\end{equation}
where $r_s$ is a measure of inter-electronic distance and $a_0$ is the Bohr radius. The value of $r_s/a_0$ ranges between 2 and 6 for typical metals. We have chosen $r_s/a_0=4$, $\varepsilon_r=10$ and $g=2$ (spin-degeneracy) which gives $C^\prime\approx0.265$ in Eq. (\ref{feggap}). The points marked by small circles in Fig.(\ref{epsilonomega}) are the plasmon energy gaps (in units of $E_F$) for the respective systems. The gaps show the following trend : $(\tilde{\omega}^{(0)}_p)_{\text{TCF}}<(\tilde{\omega}^{(0)}_p)_{\text{Weyl}}<(\tilde{\omega}^{(0)}_p)_{\text{FEG}}$. Hence, for the same set of parameters, the plasmon gap of TCFs is smaller than that of (doped) Weyl semimetal. For $E_F=10$ eV, plasmons energies will be $\approx3.3$ eV, $4.3$ eV and $4.45$ eV respectively for TCF, Weyl semimetals and FEG respectively. They may be experimentally discerned using electron energy loss spectroscopy. The peaks in Fig. (\ref{epsilonomega}) correspond to logarithmic singularities in the dielectric function. Using the small $Q$ expansion of the dynamical polarization function, the real parts of the dielectric functions for TCFs and Weyl semimetals in the long wavelength limit can be written as
\begin{equation}
\text{Re}[\varepsilon(Q\to 0,\Omega)]_\text{TCF}=1-\frac{2C}{3}\l(\frac{1}{\tilde{\omega}^2}+\log\l|\frac{1-\tilde{\omega}^2}{\Lambda^2-\tilde{\omega}^2}\r|\r)
\end{equation}
and
\begin{equation}
\text{Re}[\varepsilon(Q\to 0,\Omega)]_\text{Weyl}=1-\frac{2C}{3}\l(\frac{1}{\tilde{\omega}^2}+\frac{1}{4}\log\l|\frac{4-\tilde{\omega}^2}{4\Lambda^2-\tilde{\omega}^2}\r|\r),
\end{equation}
respectively. For TCFs, the logarithmic singularity occurs at $\tilde{\omega}=1$ i.e. $\omega=E_F/\hbar$. For Weyl fermions, it occurs at $\tilde{\omega}=2$ i.e. $\omega=2E_F/\hbar$.

The approximate plasmon dispersion in the long wavelength regime can be obtained by taking into account higher order terms of Eq.(\ref{chiexpand}). The plasmon dispersion upto the order of $Q^2$ is
\begin{equation}\label{dispersion}
\tilde{\omega}_p=\tilde{\omega}^{(0)}_p\l(1+\frac{\xi(\tilde{\omega}^{(0)}_p)C}{2\l(1+(2C/3)\log \Lambda^2\r)} Q^2 \r),
\end{equation}
where
\begin{equation}
\xi(\tilde{\omega}^{(0)}_p)=\frac{4}{15}\l(\frac{3}{2(\tilde{\omega}^{(0)}_p)^4}-\frac{1}{2(\tilde{\omega}^{(0)}_p)^2}+\frac{3}{8}\r).
\end{equation}

Thus, like other 3D electron gases, TCFs also host a gapped plasmon mode which is quadratic to lowest order in $Q$.
The plasmon mode can be traced numerically from the loss function which is defined as 
\begin{equation}\label{loss}
-\Im\l[\frac{1}{\varepsilon({\bf q},\omega)}\r]=\frac{V(q)\Im[\chi]}{\l(1-V(q)\Re[\chi]\r)^2+ \l(V(q)\Im[\chi]\r)^2}\ .
\end{equation}

Figure (\ref{lossfig}) shows the density plot of loss function for TCF. The plasmon mode can be spotted as the bright curve originating outside the PHC and finally merging into it. The part of the plasmon mode outside the PHC is undamped while that inside the PHC gets damped into particle-hole excitations, acquiring a finite lifetime. The zoomed version of the above plot is shown in Fig.(\ref{plasmonfig}), where the analytically obtained plasmon mode in Eq.(\ref{dispersion}) (labelled by dotted line) is plotted alongside the numerically obtained mode for comparison. The agreement between the two solutions holds good for low $Q$ as expected. The natural logarithm of loss function as functions of $E_F$ and photon energy $\hbar \omega$ in the limit $Q\to0$ is shown in Fig. (\ref{omegaef}). The plasmon gap (bright yellow curve) increases with $E_F$ and does not decay into PHC. The gap varies linearly with $E_F$ for higher values of $E_F$. Figure (\ref{omegaer}) shows the density plot of natural logarithm of loss function as functions of background dielectric constant $\varepsilon_r$ and frequency ($\tilde{\omega}$). As usual, the plasmon gap decreases with $\varepsilon_r$.

It is to be noted that the plasmon mode and response functions of systems with anisotropic energy dispersion are also anisotropic in nature\cite{rashbaplasmon,semidirac,sadhukhan,sdsarma,materials}. The plasmon modes of multi-Weyl fermions have also been studied recently\cite{scirep}. Multi-Weyl fermions with monopole charges 2 and 3 have similar forms of anisotropic dispersion as the quadratic $(n=2)$ and cubic $(n=3)$ TCFs, respectively. Their plasmon modes are also found to be anisotropic. So, the quadratic and cubic TCFs will also have similar nature of anisotropy in the plasmon dispersion as for multi-Weyl fermions apart from a gap renormalization due to the existence of flat band. We expect that the plasmon gap should vary as $\sim E_F^{1/n}$ and $\sim E_F$ along the directions of linear and non-linear dispersion respectively, in the long wavelength limit.

\begin{figure}
\hspace{-0.5cm}\includegraphics[trim={0cm 1cm 0cm 0.0cm},clip,width=9cm]{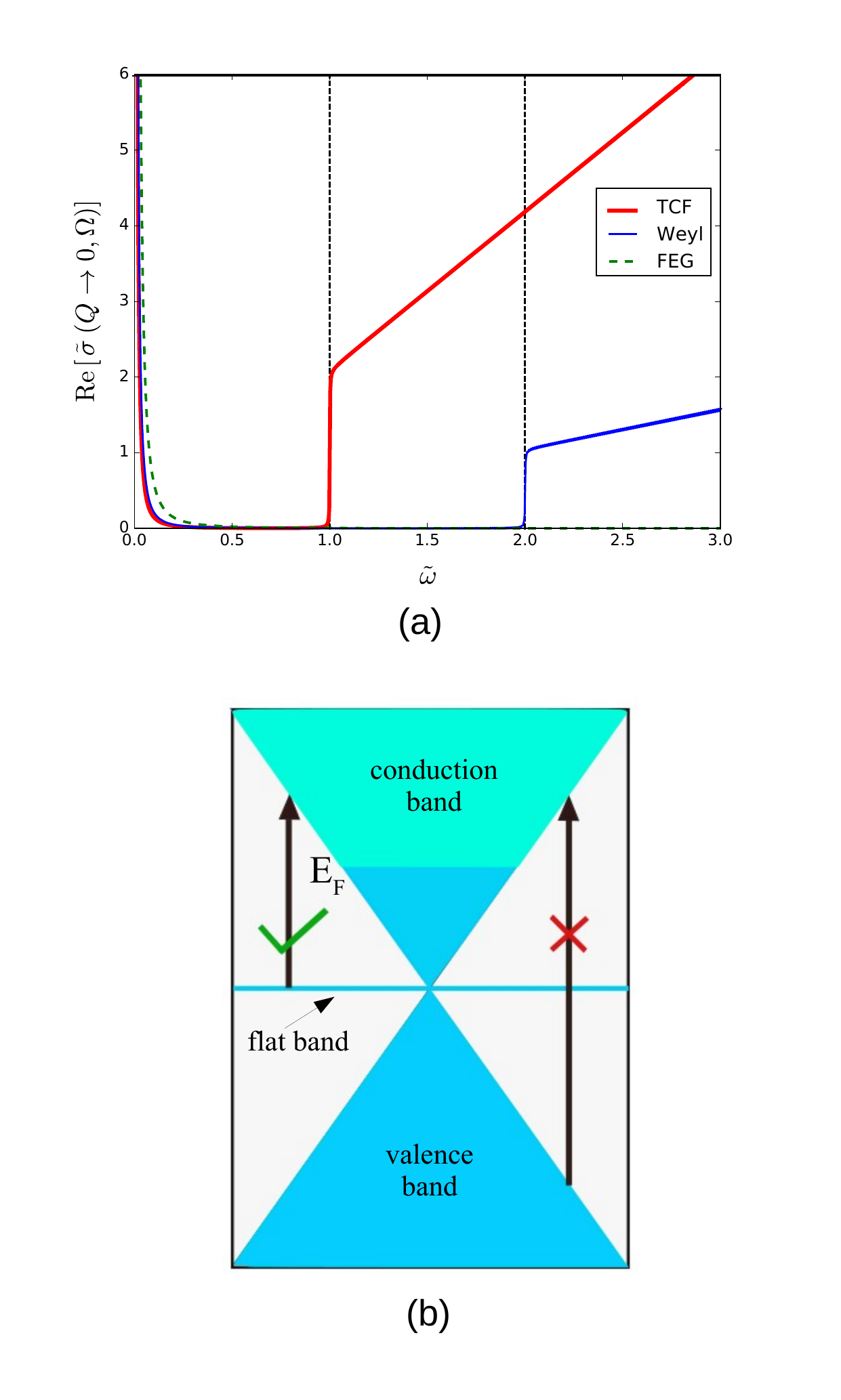}
\caption{(a) Plots of $\Re[\tilde{\sigma}(Q\to 0,\Omega)]$ vs $\tilde{\omega}$ for TCF, Weyl semimetals and FEG. The divergence at $\Omega\to0$ refers to the intraband energy absorptions of the respective systems. The optical absorption for TCFs and Weyl semimetals begin at $\tilde{\omega}=1$ and $\tilde{\omega}=2$ respectively and increase linearly with frequency. (b) Schematic diagram depicting interband transitions for TCFs in $Q\to0$ limit. The flat-to-conduction transitions are allowed while valence-to-conduction ones are highly suppressed by $Q$-dependence.}
\label{opticalnon}
\end{figure}

\begin{figure}
\hspace{-0.5cm}\includegraphics[trim={0cm 0cm 0cm 0cm},clip,width=9.5cm]{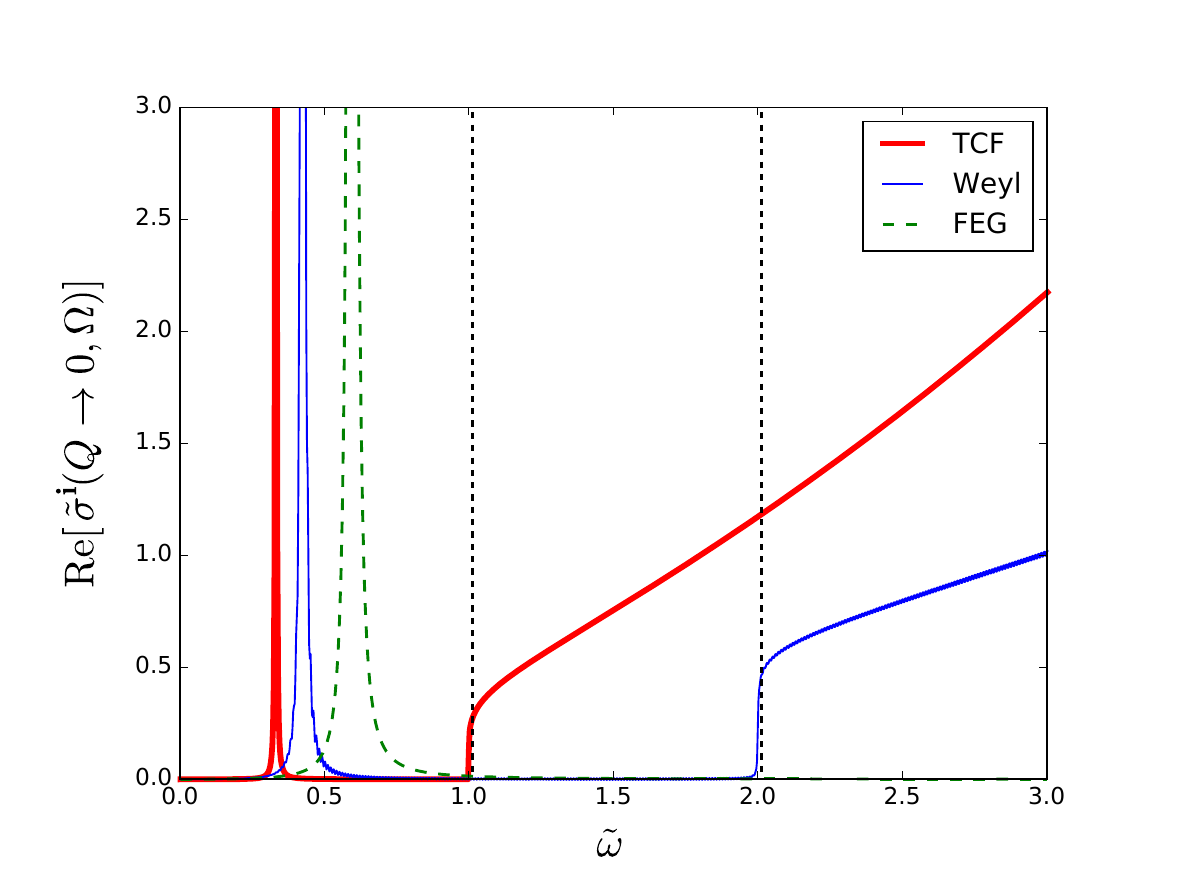}
\caption{Plots of $\Re[\tilde{\sigma}^{\bf i}(Q\to 0,\Omega)]$ vs $\tilde{\omega}$ for TCF, Weyl semimetals and FEG. Electron-electron interaction induces sharp peaks in the optical conductivities, which correspond to the plasmon modes. The optical absorption edges for TCFs and Weyl semimetals begin at $\tilde{\omega}=1$ and $\tilde{\omega}=2$ which is similar to the non-interacting case but with reduced intensities.}
\label{opticalint}
\end{figure}

\subsection{OPTICAL CONDUCTIVITY}\label{opticalsection}
The optical conductivities in the non-interacting and interacting limits are related to the respective dynamical polarization functions as\cite{flensberg}
\begin{equation}\label{opt}
\sigma({\bf q},\omega)=\frac{i\omega e^2}{q^2}\chi({\bf q},\omega)
\end{equation}
and
\begin{equation}\label{opti}
\sigma^{\bf i}({\bf q},\omega)=\frac{i\omega e^2}{q^2}\chi^{\bf i}({\bf q},\omega)
\end{equation}
respectively. The real part of optical conductivity corresponds to dissipation/absorption of energy in the medium. Using Eq.(\ref{lindqsq}) in Eqs.(\ref{opt}) and (\ref{opti}), we get
\begin{equation}
\text{Re}[\tilde{\sigma} (Q\to 0,\Omega)]=-\frac{\tilde{\omega}}{Q^2}\Im[\tilde{\chi}(Q^2,\Omega)]
\end{equation}
and
\begin{equation}\label{renorm}
\begin{aligned}
&\Re[\tilde{\sigma}^{\bf i}(Q\to 0,\Omega)]=\\& \frac{-\tilde{\omega}\ \Im[\tilde{\chi}(Q^2,\Omega)]/Q^2}{(1-C\  \Re[\tilde{\chi}(Q^2,\Omega)]/Q^2)^2+ (C\ \Im[\tilde{\chi}(Q^2,\Omega)]/Q^2 )^2},
\end{aligned}
\end{equation}
where
\begin{equation}\label{imchi}
\Im[\tilde{\chi}(Q^2,\Omega)]=-\frac{2}{3}\l[\frac{2\pi}{\tilde{\omega}} \delta(\tilde{\omega})+\pi \Theta(\tilde{\omega}^2-1)\r] Q^2
\end{equation}
and 
\begin{equation}
\Re[\tilde{\chi}(Q^2,\Omega)]=\frac{2}{3}\l(\frac{1}{\tilde{\omega}^2} - \pi^2\delta^2(\tilde{\omega})-\log\l|\frac{\Lambda^2-\tilde{\omega}^2}{1-\tilde{\omega}^2}\r|\r)Q^2.
\end{equation}
Here, we have defined $\Re[\tilde{\sigma} (Q\to 0,\omega)]=\Re[\sigma (q\to 0,\omega)]/\sigma_F$ with $\sigma_F=e^2 g k_F/(4\pi^2\hbar)$. The variation of $\Re[\tilde{\sigma} (Q\to 0,\Omega)]$ and $\Re[\tilde{\sigma}^{\bf i} (Q\to 0,\Omega)]$ with $\tilde{\omega}$ for TCFs, Weyl semimetals and 3D FEG are plotted in Figs.(\ref{opticalnon}) and (\ref{opticalint}) respectively.  The different terms of the dynamical polarization function contribute to the optical conductivity in the following way: The first term of the imaginary part of dynamical polarization function in Eq. (\ref{imchi}) is a Dirac-delta function which gives the zero frequency peak in optical conductivity, while the second term is a step function which gives the absorption edge i.e. the frequency above which particle-hole excitations occur. The zero frequency peak accounts for the intraband absorption and is evident in all the three systems. The interband absorption edges of TCFs and  Weyl semimetals commence at $\hbar\omega=E_F$ and $\hbar\omega=2E_F$ respectively and the absorption grows linearly with frequency. For TCF, the absorption edge corresponds to the onset of flat-to-conduction absorption whereas for Weyl semimetals, it indicates valence-to-conduction (or intercone) absorption. The rate of increase in interband absorption with frequency is higher in TCFs as compared to Weyl semimetals.

The intercone absorption for TCFs vanishes in the $Q\to0$ limit. The absence of intercone optical transition is a signature of pseudospin-1 Dirac/Weyl systems\cite{dora,multifold}. For two-dimensional case, it was attributed to the Berry phase of the charge carriers around the Dirac point\cite{diceoptical}. 
Here, we explain this feature for the 3D case (i.e. TCFs) by using the lowest order $Q$ dependence of the different band overlaps and $1/(\hbar \omega-\Delta E)$ factors of the dynamical polarization function (see Eq. (\ref{lindhard})) as shown in Table I. The interplay of band overlaps and $1/(\hbar\omega-\Delta E)$ factor imparts $Q^2$ dependence to the intraband and flat-to-conduction contributions but $Q^4$ dependence to the valence-to-conduction contribution in dynamical polarization function of TCFs. This makes its $\tilde{\sigma}_{vc}(Q,\Omega)\sim \mathcal{O}(Q^2)$,  which becomes vanishingly small as  $Q\to0$. However, the same valence-to-conduction contribution in Weyl semimetals is of the order of $Q^2$ for small $Q$ due to which its $\tilde{\sigma}_{vc}(Q,\Omega)$ attains a constant value as $Q\to 0$. Hence, the intercone transitions of TCFs in long wavelength limit are highly suppressed by $Q$ dependence.

\begin{table}
\begin{center}
\begin{tabular}{ | c | c | c | c | }
\hline
\textbf{Contribution} & \textbf{Band overlap} & $1/(\omega-\Delta E)$ & $\chi(Q,\Omega)$ \\
\hline
Intra-conduction & $\propto Q^0$ & $\propto Q^2$ & $\propto Q^2$ \\
\hline
Flat-to-conduction & $\propto Q^2$ & $\propto Q^0$ & $\propto Q^2$ \\
\hline
Valence-to-conduction & $\propto Q^4$ & $\propto Q^0$ & $\propto Q^4$  \\
\hline
\end{tabular}
\caption{Table for lowest order $Q$ dependence of terms in the band overlaps and $1/(\omega-\Delta E)$ factor of dynamical polarization function.}
\end{center}
\end{table}

In the interacting limit, the zero frequency peak vanishes and new peaks emerge at frequencies corresponding to the plasmon gaps. The magnitudes of interband absorption gets reduced for both Weyl fermions and TCFs but the location of the absorption edges remain unaltered. 

\section{conclusion}\label{conc}

We have explored the dynamical polarization function, static polarization function, PHC, dielectric function, plasmon mode and optical conductivity of TCFs and compared the results with those of Weyl fermions and 3D FEG. The PHC gets extended due to transitions between flat and conduction bands which occur for frequencies above $E_F/\hbar$. The static polarization function varies with momentum in a similar fashion as Weyl fermions, but has larger magnitude as
compared to the latter. The dominant contributions to the dynamical polarization function are of the order of $Q^2$ which represent intra-conduction band and flat-to-conduction transitions, while valence-to-conduction transitions are of the order of $Q^4$.  An approximate expression for the plasmon dispersion has been derived within RPA using the small $Q$ expansion of dynamical polarization function. The plasmon frequency shows the usual dependence $\omega\sim\omega_0+\omega_1 q^2$ as observed in other 3D electronic systems. Like Weyl fermions, the plasmon energy gap is proportional to $E_F$ for $E_F\gg\hbar \omega$ and is a decreasing function of background dielectric constant. The plasmon energy gap is reduced as compared to Weyl fermions for the same set of parameters and no plasmon mode occurs as $E_F\to0$.

We obtain the analytical expression of real part of optical conductivity in the $Q\to0$ limit for both nonteracting and interacting cases. Unlike Weyl semimetals, the interband optical absorption for TCFs begins at $\hbar\omega=E_F$ and the optical transitions between valence and conduction bands are highly suppressed in the long wavelength limit. We explain this feature using the lowest order $Q$ dependence of the different band overlaps and $1/(\hbar \omega-\Delta E)$ factors of the dynamical polarization function. The rate of increase in optical absorption with frequency is higher in TCFs than Weyl semimetals. On incorporating electron-electron interactions, the energy absorption gets reduced in both the systems and peaks emerge at the plasmon frequencies.

Overall, the flat band endows the response functions with several features which are in contrast to Weyl semimetals such as a new region representing flat-to-conduction absorptions in the PHC, increased static polarization, reduced plasmon energy gap with respect to Weyl fermions, shift in absorption edge, enhanced rate of
increase in optical absorption with frequency and vanishing valence-to-conduction transitions in the long wavelength
limit. The last effect demonstrates
`shielding' of the valence band by the flat band to highly suppress intercone optical transitions. These effects can be experimentally observed by optical conductivity measurements and electron energy loss spectroscopy (for
plasmon detection) which may testify the presence of TCF-like excitations in a material.

The linear response functions and plasmon dispersion may also be obtained using the lattice model of TCFs. It will not change the results of low energy model significantly as long as the Fermi energy is close to the band-touching nodes and the frequency of the perturbation is of the order of or much smaller than the Fermi energy. Under these conditions, the nature of the response functions will not alter qualitatively except for small numerical corrections arising due to higher order terms of momenta.  Also, the momentum cutoff will not be required in the lattice model and hence will not appear in the plasmon gap or dispersion.

\begin{center}
{\bf ACKNOWLEDGEMENTS}
\end{center}
We would like to thank Sonu Verma for useful discussions.

\appendix{}

\section{Theory of linear density response}\label{app}
The Hamiltonian operator of an electron gas in low energy continuum model of a lattice (excluding electron-electron interactions) is given by
\begin{equation}
\hat {H}=\sum_{{\bf k},\lambda}^{}E_{\lambda\bf k} c_{\lambda\bf k}^\dagger  c_{\lambda\bf k},
\end{equation}
where $c_{\lambda\bf k}^\dagger $ and $c_{\lambda\bf k}$ are creation and annihilation operators of the single-particle states $|\psi_{\lambda\bf k}\rangle\equiv |\lambda ({\bf k}) \rangle|{\bf k}\rangle$ with energies $E_{\lambda\bf k}$ and $\lambda$ is the band index. The density operator $\hat{\rho}({\bf r})$ is given by
\begin{equation}
\hat{\rho}({\bf r})=\hat{\Psi}^\dagger({\bf r})\hat{\Psi} ({\bf r}).
\end{equation}
The field operators $\hat{\Psi}^\dagger({\bf r})$ and $\hat{\Psi}({\bf r})$ are generally expressed in terms of operators corresponding to momentum-spin basis $\{|\psi_{s,\bf k}\rangle\}$ (i.e. $\{|s\rangle |{\bf k}\rangle\}$), which gives 
\begin{equation}\label{rhospin}
\hat{\rho}({\bf r})=\frac{1}{\mathcal{V}}\sum_{{\bf q}}^{}\e^{i {\bf q}\cdot {\bf r}}\bigg(\sum_{{\bf k},s}^{}c_{s\bf k}^\dagger c_{s{\bf k+q}}\bigg). 
\end{equation}
For a three-band system, the Hamiltonian is diagonal in $\{|\psi_{\lambda{\bf k}}\rangle\}$ basis and hence it is convenient to expand $\hat{\rho}({\bf r})$ in operators corresponding to this basis. The basis transformation equations are given by
\begin{equation}\label{transform}
c_{s {\bf k} }=\sum_{\lambda}^{}\langle s|\lambda({\bf k}) \rangle c_{\lambda{\bf k} }, \hspace{0.3cm}
c^\dagger_{s {\bf k}}=\sum_{\lambda}^{}\langle s|\lambda({\bf k}) \rangle^* c^\dagger_{\lambda {\bf k}},
\end{equation}
where $\lambda$ is summed over $(-1,0,1)$. Using Eqs.(\ref{rhospin}) and (\ref{transform}), we get
\begin{equation}
\hat{\rho}({\bf r})=\frac{1}{\mathcal{V}}\sum_{{\bf q}}^{}\e^{i {\bf q}\cdot {\bf r}}\bigg(\sum_{{\bf k},\lambda_1,\lambda_2}^{}\langle \lambda_1 ({\bf k}) | \lambda_2 ({\bf k+q}) \rangle c_{\lambda_1{\bf k} }^\dagger c_{\lambda_2{\bf k+q}}\bigg).
\end{equation}
When the system is in thermodynamic equilibrium with a reservoir at temperature $T$, the equilibrium electron density $\rho({\bf r})$ given by 
\begin{equation}
\rho({\bf r})\equiv\langle \hat{\rho} ({\bf r}) \rangle_0= \frac{1}{Z_0}\sum_{\{N\}}\langle N |\hat{\rho} ({\bf r}) \e^{-\beta \hat {H}} | N \rangle,
\end{equation}
where $Z_0=\sum_{\{N\}}^{}\langle N | \e^{-\beta \hat {H}} | N \rangle$ is the canonical partition function, $\beta=(k_B T)^{-1}$ and the summation runs over all the $N$-particle fermionic eigenstates of $\hat{H}$. 
When the system is subjected to an external electric field ${\bf E}_{\text{ext}}({\bf r},t)$, a perturbation of the form
\begin{equation}
\hat{V}(t)=\int \hat{\rho}({\bf r}^\prime) \phi_{\text{ext}}({\bf r}^\prime,t) d{\bf r}^\prime \Theta (t-t_0)
\end{equation}
gets added to the Hamiltonian $\hat{H}$, where $\phi_{\text{ext}}({\bf r}^\prime,t)=-e\int^{{\bf r}^\prime} {\bf E}_{\text{ext}}({\bf r},t)\cdot {\bf r} \ d{\bf r}$ is the electric potential and $t_0$ is the time when the field is switched on. The new Hamiltonian is
\begin{equation}
\hat{H}^\prime(t)=\hat{H}+\hat{V}(t).
\end{equation}
The time evolution of the states are now governed by $\hat{H}^\prime(t)$, which drives the system out of equilibrium and the electron density becomes a function of both space and time in general. Considering magnitude of the perturbation very small compared to $\langle \hat{H} \rangle_0$, the nonequilibrium expectation value of density upto linear order in $\phi_{\text{ext}}$ is given by the Kubo formula as\cite{flensberg}
\begin{equation}
\langle \hat{\rho} ({\bf r}) \rangle= \langle \hat{\rho} ({\bf r}) \rangle_0+\int d{\bf r}^\prime \int_{t_0}^{\infty}  dt^\prime \chi({\bf r},{\bf r^\prime},t,t^\prime) \phi_{\text{ext}}({\bf r}^\prime,t^\prime)
\end{equation}
or,
\begin{equation}\label{rhoind}
\rho_{\text{ind}}({\bf r},t)=\int d{\bf r}^\prime \int_{t_0}^{\infty}  dt^\prime \chi({\bf r},{\bf r^\prime},t,t^\prime) \phi_{\text{ext}}({\bf r}^\prime,t^\prime),
\end{equation}
where $\rho_{\text{ind}}({\bf r},t)\equiv \langle \hat{\rho} ({\bf r}) \rangle - \langle \hat{\rho} ({\bf r}) \rangle_0$ is the induced density and $\chi({\bf r},{\bf r^\prime},t-t^\prime)$ is the retarded density-density correlation function or polarizability given by
\begin{equation}
\chi({\bf r},{\bf r^\prime},t,t^\prime)=-i\Theta(t-t^\prime)\langle[\hat{\rho}_{\text{I}}({\bf r},t),\hat{\rho}_{\text{I}}({\bf r}^\prime,t^\prime)]\rangle_0/\hbar.
\end{equation}
Here, $\langle...\rangle_0$ denotes the expection value taken with respect to the equilibrium state and $\hat{\rho}_I({\bf r},t)$ is the density operator in the interaction picture, which is defined as
\begin{equation}
\hat{\rho}_{\text{I}}({\bf r},t)=\e^{i \hat{H}t/\hbar} \hat{\rho}({\bf r})\e^{-i \hat{H}t/\hbar}.
\end{equation}
It can be seen that the polarizability is non-local in space and retarded in time, i.e. the response at a particular point in space at a given instant of time is correlated to the value of external field at some other point in space at any previous instant of time.
Moreover, $\langle[\hat{\rho}_{\text{I}}({\bf r},t),\hat{\rho}_{\text{I}}({\bf r}^\prime,t^\prime)]\rangle_0$ is always a function of $(t-t^\prime)$ and for translationally invariant systems, it is a function of ${\bf r}-{\bf r^\prime}$. For such systems, $\chi({\bf r},{\bf r^\prime},t,t^\prime)\equiv\chi({\bf r}-{\bf r^\prime},t-t^\prime)$ and hence $\rho_{\text{ind}}({\bf r},t)$ becomes the convolution of $\chi$ and $\phi_{\text{ext}}$ in both time and space coordinates. By convolution theorem, we get
\begin{equation}
\rho_{\text{ind}}({\bf q},\omega)= \chi({\bf q},\omega) \phi_{\text{ext}}({\bf q},\omega),
\end{equation}
where 
\begin{equation}\label{lind1}
\begin{aligned}
\chi({\bf q},\omega)=\int d({\bf r}-{\bf r}^\prime)& \int_{-\infty}^{\infty} d(t-t^\prime) \chi({\bf r}-{\bf r^\prime},t-t^\prime)\times \\& \e^{-i[{\bf q}\cdot ({\bf r}-{\bf r}^\prime)-\omega (t-t^\prime)]}
\end{aligned}
\end{equation}
and
\begin{equation}
\phi_{\text{ext}}({\bf q},\omega)=\int d{\bf r}^\prime \int_{-\infty}^{\infty} d t^\prime \phi_{\text{ext}}({\bf r^\prime},t^\prime) \e^{-i({\bf q}\cdot {\bf r}^\prime-\omega t^\prime) }
\end{equation}
are the Fourier transforms.
On simplication, Eq.(\ref{lind1}) reduces to
\begin{equation}\label{lindhardnon}
\chi({\bf q}, \omega)=\lim_{\eta\to 0}\frac{g}{V}\sum_{{\bf k},\lambda,\lambda^\prime}
\frac{F_{\lambda,\lambda^\prime}({\bf k,k+q}) (f_{\lambda,{\bf k}} - f_{\lambda^\prime,{\bf k+q}})}{\hbar (\omega + i \eta) + E_{\lambda,{\bf k}}- E_{\lambda^\prime,{\bf k+q}}}.
\end{equation} 
This is called the dynamical polarization function. In (\ref{lindhardnon}), $g$ is the degeneracy factor, $F_{\lambda,\lambda^\prime}({\bf k,k+q})=|\langle \lambda({\bf k})|\lambda^\prime({\bf k+q})\rangle|^2$ is the overlap between the corresponding states and $f_{\lambda,{\bf k}}=[\e^{\beta(E_{\lambda,{\bf k}}-E_F)}+1]^{-1}$ is the Fermi-Dirac distribution function. 

On incorporating electron-electron interactions, the dynamical polarization function obtained within Random Phase Approximation (RPA) is given by
\begin{equation}\label{lindint}
\chi^{\bf i}({\bf q},\omega)=\frac{\chi({\bf q},\omega)}{\varepsilon({\bf q},\omega)},
\end{equation}
where superscript ${\bf i}$ stands for `interactions', $\chi({\bf q},\omega)$ is the non-interacting dynamical polarization function given by Eq.(\ref{lindhardnon}), and $\varepsilon({\bf q},\omega)$ is the dielectric function which has the following form :
\begin{equation} {\label{diel}}
\varepsilon({\bf q},\omega)=1-V(q) \chi ({\bf q},\omega).
\end{equation}
Here $V(q)=e^2/(\varepsilon_r \varepsilon_0 q^2)$ is the Fourier transform of Coulomb potential energy between electrons in SI units in a medium of background dielectric constant $\varepsilon_r$. The real space-time dielectric function $\varepsilon({\bf r},t)$ is the inverse Fourier transform of Eq.(\ref{diel}) and acts as a response function between $\phi_{\text{ext}}$ and $\phi_{\text{total}}$:
\begin{equation}
\phi_{\text{ext}}({\bf r},t)=\int d{\bf r}^\prime \int_{t_0}^{\infty}  dt^\prime \varepsilon({\bf r}-{\bf r^\prime},t-t^\prime) \phi_{\text{total}}({\bf r}^\prime,t^\prime).
\end{equation}
The poles of the interacting dynamical polarization function in Eq.(\ref{lindint}) or the zeroes of the dielectric function in  Eq.(\ref{diel}) correspond to the collective modes of electron oscillations and are known as plasmon modes. They can be damped or undamped depending on the values of $Q$ and $\Omega$ of the external perturbation. The undamped plasmon modes $\Omega_p$ are obtained from the zeroes of ${\Re [\varepsilon({\bf q},\omega)}]$ in the region where $\Im[\chi ({\bf q},\omega)]$ vanishes.\\

\section{Alternative derivation of real part of optical conductivity}

In long wavelength limit $(q\to0)$, $\Re[\sigma_{xx} (\omega)]$ (excluding the zero frequency peak) can be analytically derived from Kubo formula as
\begin{equation}
\begin{aligned}
&\Re[\sigma_{xx} (\omega)]=
\\&\frac{\pi g e^2}{ (2\pi)^d \omega}\sum_{\lambda,\lambda^\prime}^{} \int d^d {\bf k} (f_{\lambda^\prime}({\bf k})-f_{\lambda}({\bf k}))|v^{\lambda^\prime\lambda}_x|^2\delta(\Delta E_{\lambda\lambda^\prime}-\hbar \omega),
\end{aligned}
\end{equation}
where $\Delta E_{\lambda\lambda^\prime}=E_\lambda({\bf k})-E_{\lambda^\prime}({\bf k})$, $d$ is the dimensionality, $g$ is the degeneracy and $v^{\lambda^\prime\lambda}_x=\langle\psi_{\lambda^\prime}({\bf k})|\hat{v}_x|\psi_{\lambda}({\bf k})\rangle$ with $\hat{v}_x=\partial_{k_x}H/\hbar$ being the $x$-component of velocity operator. For TCFs, the above expression reduces to
\begin{equation} \label{optical-tcf-exp}
\begin{aligned}
&\Re[\sigma_{xx} (\omega)]=
\\&\frac{g e^2}{ 8 \pi^2 \omega} \int d^3 {\bf k} \bigg[(f_{-}(k)-f_{+}(k))|v^{-+}_x|^2\delta(2\hbar v_F k-\hbar \omega)\\&
+(f_{0}(k)-f_{+}(k))|v^{0+}_x|^2\delta(\hbar v_F k-\hbar \omega)\bigg].
\end{aligned}
\end{equation}
For TCF, $\hat{v}_x=v_F S_x$, $v^{-+}_x=0$ and $|v^{0+}_x|^2=v^2_F(3-\cos 2\phi+2 \cos^2\phi \cos 2\theta)/8$. Using these results, Eq.(\ref{optical-tcf-exp}) gives
\begin{equation}
\Re[\sigma_{xx} (\omega)]=\frac{g e^2 \omega}{6 \pi \hbar v_F} \Theta(\omega-v_F k_F),
\end{equation}
where $\Theta(x)$ is the usual step function. Unlike Weyl semimetals, $v^{-+}_x=0$ in TCF. This feature is also seen in dice lattice, where it was attributed to zero (modulo $2\pi$) Berry phase of the charge carriers\cite{diceoptical}.


\begin{thebibliography}{55}


\bibitem{herring}
C. Herring, Phys. Rev. {\bf 52}, 365 (1937).

\bibitem{wan-weyl}
X. Wan, A. M. Turner, A. Vishwanath, and S. Y. Savrasov, Phys. Rev. B {\bf 83}, 205101 (2011).

\bibitem{xu}
G. Xu, H. Weng, Z. Wang, X. Dai, and Z. Fang, Phys. Rev. Lett. {\bf 107}, 186806 (2011).

\bibitem{burkov}
A. A. Burkov, and L. Balents, Phys. Rev. Lett. {\bf 107}, 127205 (2011).

\bibitem{bulmash}
D. Bulmash, C.-X. Liu, and X.-L. Qi, Phys. Rev. B {\bf 89}, 081106 (2014).

\bibitem{murakami}
S. Murakami, New J. Phys. {\bf 9}, 356 (2007).

\bibitem{halasz}
G. B. Halász, and L. Balents, Phys. Rev. B {\bf 85}, 035103 (2012).

\bibitem{smhuang}
S.-M. Huang, S.-Y. Xu, I. Belopolski, C.-C. Lee, G. Chang, B. Wang, N. Alidoust, G. Bian, M. Neupane, C. Zhang, S. Jia, A. Bansil, H. Lin, and M. Z. Hasan, Nat. Commun. {\bf 6}, 7373 (2015).

\bibitem{lvb}
B. Q. Lv, H. M. Weng, B. B. Fu, X. P. Wang, H. Miao, J. Ma, P. Richard, X. C. Huang, L. X. Zhao, G. F. Chen, Z. Fang, X. Dai, T. Qian, and H. Ding,, Phys. Rev. X {\bf 5}, 031013 (2015).

\bibitem{lvc}
B. Q. Lv, N. Xu, H. M. Weng, J. Z. Ma, P. Richard, X. C. Huang, L. X. Zhao, G. F. Chen, C. E. Matt, F. Bisti, V. N. Strocov, J. Mesot, Z. Fang, X. Dai, T. Qian, M. Shi, and H. Ding, Nat. Phys. {\bf 11}, 724 (2015).

\bibitem{wengfang}
H. Weng, C. Fang, Z. Fang, B. A. Bernevig, and X. Dai, Phys. Rev. X {\bf 5}, 011029 (2015).

\bibitem{xub}
S.-Y. Xu, I. Belopolski, N. Alidoust, M. Neupane, G. Bian, C. Zhang, R. Sankar, G. Chang, Z. Yuan, C.-C. Lee, S.-M. Huang, H. Zheng, J. Ma, D. S. Sanchez, B. Wang, A. Bansil, F. Chou, P. P. Shibayev, H. Lin, S. Jia, and  M. Z. Hasan , Science {\bf 349}, 613 (2015).

\bibitem{abrikosov}
A. Abrikosov, and S. Beneslavskii, Sov. Phys. JETP {\bf 32}, 699 (1971).

\bibitem{wangdirac}
Z. Wang, Y. Sun, X.-Q. Chen, C. Franchini, G. Xu, H. Weng, X. Dai, and Z. Fang, Phys. Rev. B {\bf 85}, 195320 (2012).

\bibitem{young}
S. M. Young, S. Zaheer, J. C. Y. Teo, C. L. Kane, E. J. Mele, and A. M. Rappe, Phys. Rev. Lett. {\bf 108}, 140405 (2012).

\bibitem{murakamietal}
S. Murakami, S. Iso, Y. Avishai, M. Onoda, and N. Nagaosa, Phys. Rev. B {\bf 76}, 205304 (2007).

\bibitem{steinberg}
Steinberg, J. A., S. M. Young, S. Zaheer, C. L. Kane, E. J. Mele, and A. M. Rappe, Phys. Rev. Lett. {\bf 112}, 036403 (2014).

\bibitem{borisenko}
S. Borisenko, Q. Gibson, D. Evtushinsky, V. Zabolotnyy, B. Buechner, and R. J. Cava, Phys. Rev. Lett. {\bf 113}, 027603 (2014).

\bibitem{liua}
Z. K. Liu, J. Jiang, B. Zhou, Z. J. Wang, Y. Zhang, H. M. Weng, D. Prabhakaran, S-K. Mo, H. Peng, P. Dudin, T. Kim, M. Hoesch, Z. Fang, X. Dai, Z. X. Shen, D. L. Feng, Z. Hussain, and Y. L. Chen, Nat. Mater. {\bf 13}, 677 (2014).

\bibitem{liub}
Z. K. Liu, B. Zhou, Y. Zhang, Z. J. Wang, H. M. Weng, D. Prabhakaran, S.-K. Mo, Z. X. Shen, Z. Fang, X. Dai, Z. Hussain, and Y. L. Chen, Science {\bf 343}, 864 (2014).

\bibitem{neupane}
M. Neupane, S.-Y. Xu, R. Sankar, N. Alidoust, G. Bian, C. Liu, I. Belopolski, T.-R. Chang, H.-T. Jeng, H. Lin, A. Bansil, F. Chou, and M. Z. Hasan, Nat. Commun. {\bf 5}, 3786 (2014).

\bibitem{adler}
S. L. Adler, Phys. Rev. {\bf 177}, 2426 (1969).

\bibitem{bell}
J. S. Bell, and R. W. Jackiw, Nuovo Cimento {\bf 60}, 47 (1969).

\bibitem{nielson}
H. B. Nielsen, and M. Ninomiya, Phys. Lett. B {\bf 130}, 389 (1983).


\bibitem{yang}
K.-Y. Yang, Y.-M. Lu, and Y. Ran, Phys. Rev. B {\bf 84}, 075129 (2011).

\bibitem{weyl}
H. Weyl, Proc. Natl. Acad. Sci. U.S.A. {\bf 15}, 323 (1929).

\bibitem{dirac}
P. A. M. Dirac, Proc. R. Soc. A {\bf 117}, 610 (1928).


\bibitem{bradlyn}
B. Bradlyn, J. Cano, Z. Wang, M. Vergniory, C. Felser, R. Cava, and B. A. Bernevig, Science {\bf 353}, 496 (2016).


\bibitem{solu}
A. A. Soluyanov, D. Gresch, Z. Wang, Q. Wu, M. Troyer, X. Dai, and B. A. Bernevig, Nature (London) {\bf 527}, 495 (2015).

\bibitem{xuzhang}
Y. Xu, F. Zhang, and C. Zhang, Phys. Rev. Lett. {\bf 115}, 265304 (2015).

\bibitem{wieder}
B. J. Wieder, Y. Kim, A. M. Rappe, and C. L. Kane, Phys. Rev. Lett. {\bf 116}, 186402 (2016).

\bibitem{chang2}
G. Chang, S.-Y. Xu, S.-M. Huang, D. S. Sanchez, C.-H. Hsu, G. Bian, Z.-M. Yu, I. Belopolski, N. Alidoust, H. Zheng, T.-R. Chang, H.-T. Jeng, S. A. Yang, T. Neupert, H. Lin, and M. Z. Hasan, Scientific Reports {\bf 7}, 1688 (2017).

\bibitem{fulga}
I. C. Fulga and A. Stern, Phys. Rev. B {\bf 95}, 241116 (2017).

\bibitem{weng1}
H. Weng, C. Fang, Z. Fang, and X. Dai, Phys. Rev. B {\bf 94}, 165201 (2016).

\bibitem{weng2}
H. Weng, C. Fang, Z. Fang, and X. Dai, Phys. Rev. B {\bf 93}, 241202 (2016).

\bibitem{cheung}
C.-H. Cheung, R. C. Xiao,M.-C. Hsu, H.-R. Fuh, Y.-C. Lin, and C.-R. Chang, arXiv:1709.07763.

\bibitem{li}
J. Li, Q. Xie, S. Ullah, R. Li, H. Ma, D. Li, Y. Li, and X.-Q. Chen, Phys. Rev. B {\bf 97}, 054305 (2018).

\bibitem{bqlv}
B. Q. Lv, Z.-L. Feng, Q.-N. Xu, X. Gao, J.-Z. Ma, L.-Y. Kong, P. Richard, Y.-B. Huang, V. N. Strocov, C. Fang, H.-M. Weng, Y.-G. Shi, T. Qian, and H. Ding, Nature (London) {\bf 546}, 627 (2017).

\bibitem{he}
J. B. He, D. Chen, W. L. Zhu, S. Zhang, L. X. Zhao, Z. A. Ren, and G. F. Chen, Phys. Rev. B {\bf 95}, 195165 (2017).

\bibitem{chang3}
G. Chang, S.-Y. Xu, B. J. Wieder, D. S. Sanchez, S.-M. Huang, I. Belopolski, T.-R. Chang, S. Zhang, A. Bansil, H. Lin, and M. Z. Hasan, Phys. Rev. Lett. {\bf 119}, 206401 (2017).

\bibitem{tang}
P. Tang, Q. Zhou, and S.-C. Zhang, Phys. Rev. Lett. {\bf 119}, 206402 (2017).

\bibitem{zhu}
Z. Zhu, G. W. Winkler, Q. S. Wu, J. Li, and A. A. Soluyanov, Phys. Rev. X {\bf 6}, 031003 (2016).

\bibitem{takane}
D. Takane, Z. Wang, S. Souma, K. Nakayama, T. Nakamura, H. Oinuma, Y. Nakata, H. Iwasawa, C. Cacho, T. Kim, K. Horiba, H. Kumigashira, T. Takahashi, Y. Ando, and T. Sato,  Phys. Rev. Lett. {\bf 122}, 076402 (2019).

\bibitem{rao}
Z. Rao, H. Li, T. Zhang, S. Tian, C. Li, B. Fu, C. Tang,
L. Wang, Z. Li, W. Fan, J. Li, Y. Huang, Z. Liu, Y. Long,
C. Fang, H. Weng, Y. Shi, H. Lei, Y. Sun, T. Qian, and
H. Ding, Nature {\bf 567}, 496 (2019).

\bibitem{sanchez}
D. S. Sanchez, I. Belopolski, T. A. Cochran, X. Xu, J.- X. Yin, G. Chang, W. Xie, K. Manna, V. Su, C.-Y.
Huang, N. Alidoust, D. Multer, S. S. Zhang, N. Shumiya, X. Wang, G.-Q. Wang, T.-R. Chang, C. Felser,
S.-Y. Xu, S. Jia, H. Lin, and M. Z. Hasan, Nature {\bf 567}, 500 (2019).

\bibitem{lvreview}
B. Q. Lv, T. Qian, and H. Ding, Rev. Mod. Phys. {\bf 93}, 025002 (2021).

\bibitem{bitan}
S. Nandy, S. Manna, D. Calugaru, and B. Roy, Phys. Rev. B {\bf 100}, 235201 (2019).

\bibitem{bradlyn2}
B. Bradlyn, L. Elcoro, J. Cano, M. G. Vergniory, Z. Wang, C. Felser, M. I. Aroyo, and B. A. Bernevig, Topological quantum chemistry, Nature (London) {\bf 547}, 298 (2017).

\bibitem{chang}
G. Chang, B. J. Wieder, F. Schindler, D. S. Sanchez, I. Belopolski, S.-M. Huang, B. Singh, D. Wu, T.-R. Chang, T. Neupert, S.-Y. Xu, H. Lin, and M. Z. Hasan, Nat. Mater. {\bf 17}, 978 (2018).





\bibitem{flensberg}
H. Bruus and K. Flensberg, {\it Many-Body Quantum Theory in Condensed Matter Physics: An Introduction} (Oxford University Press, Oxford, UK, 2004)

\bibitem{kubo}
R. Kubo, Journal of the Physical Society of Japan, {\bf 12}, 570 (1957).

\bibitem{lindhard}
J. Lindhard, {\it Danske Matematisk-fysiske Meddeleiser}, {\bf 28} (8): 1–57 (1954). 

\bibitem{giuliani}
G. Giuliani and G. Vignale, {\it  Quantum Theory of the Electron Liquid} (Cambridge University Press, Cambridge, UK, 2005).




\bibitem{rpa1}
D. Bohm and D. Pines, Phys. Rev. {\bf 82}, 625 (1951).

\bibitem{rpa2}
D. Pines and D. Bohm, Phys. Rev. {\bf 85}, 338 (1952).

\bibitem{rpa3}
D. Bohm and D. Pines, Phys. Rev. {\bf 92}, 609 (1953).

\bibitem{plasmonics1}
E. Ozbay, Science {\bf 311}, 189 (2006).

\bibitem{plasmonics2}
N. Meinzer, W. L. Barnes and I. R. Hooper, Nat. Photon. {\bf 8}, 889 (2014).

\bibitem{plasmonics3}
D. C. Marinica, M. Zatapa, P. Nordlander, A. K. Kazansky, P. M. 
Echenique, J. Aizpurua and A. G. Borisov, Sci. Adv. {\bf 1}, e1501095 (2015).

\bibitem{stern}
F. Stern, Phys. Rev. Lett. {\bf 18}, 546 (1967).


\bibitem{gonzalez}
J. Gonzalez, F. Guinea, and M. A. H. Vozmediano, Nucl. Phys. {\bf 424}, 595 (1994).

\bibitem{wangchakra}
X. F. Wang and T. Chakraborty, Phys. Rev. B {\bf 75}, 033408 (2007); {\bf 75}, 041404 (R) (2007).

\bibitem{sarma}
E. H. Hwang and S. D. Sarma, Phys. Rev. B {\bf 75}, 205418 (2007).

\bibitem{shung}
K. W.-K. Shung, Phys. Rev. B {\bf 34}, 979 (1986).

\bibitem{ando}
T. Ando, J. Phys. Soc. Jpn. {\bf 75}, 074716 (2006).

\bibitem{wunsh}
B. Wunsch, T. Stauber, F. Sols, and F. Guinea, New J. Phys. {\bf 8}, 318 (2006).

\bibitem{pyat}
P. K. Pyatkovskiy, J. Phys.: Condens. Matter {\bf 21}, 025506 (2009).

\bibitem{diceplasmon}
J. D. Malcolm and E. J. Nicol, Phys. Rev. B {\bf 93}, 165433 (2016).

\bibitem{mahan}
G. D. Mahan, {\it Many Particle Physics} (Plenum, New York,1993).

\bibitem{sonu}
S. Verma, A. Kundu, and T. K. Ghosh, Phys. Rev. B {\bf 102}, 195208 (2020).

\bibitem{minlv}
M. Lv and S.-C. Zhang, Int. J. Mod. Phys. B  {\bf 27}, 1350177 (2013).

\bibitem{zhou}
J. Zhou, H.-R. Chang, and D. Xiao, Phys. Rev. B {\bf 91}, 035114 (2015).

\bibitem{sdhwang}
S. Das Sarma and E. H. Hwang, Phys. Rev. Lett. {\bf 102}, 206412(2009).

\bibitem{massivedirac1}
R. Sachdeva, A. Thakur, G. Vignale and A. Agarwal, Phys. Rev. B {\bf 91}, 205426 (2015).

\bibitem{massivedirac2}
A. Thakur, R. Sachdeva and A. Agarwal, J. Phys.: Condens. Matter {\bf 29}, 105701 (2017).

\bibitem{amit-prb}
A. Thakur, K. Sadhukhan, and A. Agarwal,
Phys. Rev. {\bf B} 97, 035403 (2018).

\bibitem{plasmonexp}
L. Marton, J. L. Simpson, H. A. Fowler, and N. Swanson, Phys. Rev. {\bf 126}, 182 (1962).

\bibitem{graphene-ando}
T. Ando, Y. Zheng and H. Suzuura, J. Phys. Soc. Japan {\bf 71}, 1318 (2002).

\bibitem{Gusysin}
V. P. Gusynin, S. G. Sharapov, and J. P. Carbotte, Phys. Rev. Lett. {\bf 96}, 256802 (2006). 

\bibitem{nair}
R. R. Nair, P. Blake,  A. N. Grigorenko, K. S. Novoselov, T. J. Booth, T. Stauber,
N. M. R. Peres and A. K. Geim, Science {\bf 320}, 1308 (2008).

\bibitem{mak}
K. F. Mak, M. Y. Sfeir, Y. Wu, C. H. Lui, J. A. Misewich, and
T. F. Heinz, Phys. Rev. Lett. {\bf 101}, 196405 (2008).

\bibitem{stauber}
T. Stauber, N. M. R. Peres, and A. K. Geim, Phys. Rev. B {\bf 78}, 085432 (2008).

\bibitem{dora}
B. Dóra, J. Kailasvuori, and R. Moessner, Phys. Rev. B {\bf 84}, 195422 (2011).

\bibitem{diceoptical}
E. Illes, J. P. Carbotte, and E. J. Nicol, Phys. Rev. B {\bf 92}, 245410 (2015).

\bibitem{hosur}
P. Hosur, S. A. Parameswaran, and A. Vishwanath, Phys. Rev. Lett. {\bf 108}, 046602
(2012).

\bibitem{bacsi}
A. Bacsi and A. Virozstek, Phys. Rev. B {\bf 87}, 125425 (2013).

\bibitem{ashby}
P. E. C. Ashby and J. P. Carbotte, Phys. Rev. B {\bf 89}, 245121 (2014).

\bibitem{multifold}
M.-A. S.-Martinez, F. D. Juan, and A. G. Grushin, Phys. Rev. B {\bf 99}, 155145 (2019).


\bibitem{scirep}
S. Ahn, E. Hwang and H. Min, Sci. Rep. {\bf 6}, 34023 (2016).

\bibitem{rashbaplasmon}
S. M. Badalyan, A. M.- Abiague, G. Vignale, and J. Fabian,
Phys. Rev. B {\bf 79}, 205305 (2009).

\bibitem{semidirac}
S. Banerjee and W. E. Pickett, Phys. Rev B {\bf 86}, 075124 (2012).

\bibitem{sadhukhan}
K. Sadhukhan and A. Agarwal,
Phys. Rev. B {\bf 96}, 035410 (2017).

\bibitem{sdsarma}
S. Ahn and S. D. Sarma, Phys. Rev B {\bf 103}, L041303 (2021).

\bibitem{materials}
R. Hayn, T. Wei, V. M. Silkin, and J. V. D Brink,
Phys. Rev. Materials {\bf 5}, 024201 (2021).



\end{thebibliography}
\end{document}